\documentclass[12pt]{iopart}
\usepackage{xcolor}
\usepackage{hyperref}
\usepackage{braket}
\usepackage{amsfonts}
\expandafter\let\csname equation*\endcsname\relax
\expandafter\let\csname endequation*\endcsname\relax
\usepackage{amsmath}
\usepackage{graphicx}
\usepackage{comment}
\usepackage{mathtools}
\usepackage{cite}
\usepackage{enumitem}
\usepackage{bm}
\usepackage{array}

\def\Q{\mathcal{Q}}

\begin{document}

\title{Fluctuation Theorems for Heat exchanges between passive and active baths}

\author{Massimiliano Semeraro$^{1*}$, Antonio Suma$^1$ and Giuseppe Negro$^1$}

\address{$^1$Dipartimento di Fisica, Universit\`a degli Studi di Bari and INFN, Sezione di Bari, via Amendola 173, 70126 Bari, Italy}
\ead{* \href{mailto:massimiliano.semeraro@uniba.it}{massimiliano.semeraro@uniba.it}}
\vspace{10pt}

\begin{abstract} 
In addition to providing general constraints on probability distributions, fluctuation theorems allow to infer essential information on the role played by temperature in heat exchange phenomena. In this numerical study, we measure the temperature of an out of equilibrium active bath using a fluctuation theorem that relates the fluctuations of the heat exchanged between two baths to their temperatures. Our setup consists of a single particle moving between two wells of a quartic potential accommodating two different  baths. The heat exchanged between the two baths is monitored according to two definitions: as the kinetic energy carried by the particle whenever it jumps from one well to the other and as the work performed by the particle on one of the two baths when immersed in it. First, we consider two equilibrium baths at two different temperatures and verify that a fluctuation theorem featuring the baths temperatures holds for both heat definitions. Then, we introduce an additional Gaussian coloured noise in one of the baths, so as to make it effectively an active (out-of-equilibrium) bath. We find that a fluctuation theorem is still satisfied with both heat definitions. Interestingly, in this case the temperature obtained through the fluctuation theorem for the active bath corresponds to the kinetic temperature when considering the first heat definition, while it is larger with the second one. We interpret these results by looking at the particle jump phenomenology.
\end{abstract}
\noindent{\it Keywords\/}: heat exchange, out-of-equilibrium systems, fluctuation theorem, active bath, out-of-equilibrium temperatures

\section{Introduction}
\label{sec:intro}

A fundamental open issue in statistical physics is the extension of the equilibrium framework to out of equilibrium settings. Amongst the many questions still waiting for an answer, the definition of a proper temperature that consistently regulates {\it heat fluctuations and exchanges} between out of equilibrium thermal baths posits a central problem. For glassy systems, whose non-equilibrium character is due to very long relaxational times \cite{cugliandolo1997, cugliandolo2000}, an effective thermal picture has already emerged. Here an effective temperature can in fact be defined using the non-equilibrium deviations of the fluctuation-dissipation theorem \cite{cugliandolo2011, Ilg2006}. One then naturally wonders if a similar scenario applies also to other classes out of equilibrium systems.

One class of out of equilibrium systems that in the past few years attracted great interest is {\it active matter} \cite{ramaswamy2010a,romanczuk2012, marchetti2013, elgeti2015, bechinger2016, fodor2018, carenza2019, negro2019, gompper2020, carenza2020, favuzzi2021, giordano2021, head2024}. The distinctive feature of all systems from this class is a continuous conversion and injection of energy from internal reservoirs or the surrounding environment into the system itself to produce self propulsion of its minimal constituents. Interestingly, the mere introduction of a self-propulsion mechanism results in a wealth of new phenomena and features, as for example collective motion \cite{vicsek2012, grandpre2018, gompper2020, negro2022, caporusso2024}, motility-induced phase separation \cite{tailleur2008, cates2015, caporusso2020} a rich phase diagram \cite{fily2014, cugliandolo2017, digregorio2018, petrelli2018} and dynamical phase transitions \cite{cagnetta2017, gradenigo2019, semeraro2023a}, most of which have no equivalent in passive counterparts. According to stochastic thermodynamics \cite{seifert2012, peliti2021, shiraishi2023}, the injection of energy is an irreversible process which makes active systems inherently out of equilibrium \cite{fodor2021, byrne2022}. As a consequence, the Stokes-Einstein relation between injection and dissipation of energy is naturally violated at microscopic scales \cite{burkholder2019, caprini2021b, dalcengio2021}, therefore making active matter systems a perfect stage for introducing and testing different definitions of out of equilibrium temperatures. In this respect, we mention that there have been several attempts to describe this inherent non-equilibrium character at macroscopic scales through the introduction of an effective temperature \cite{loi2008, cugliandolo2008, gov2019, palacci2010, suma2014, szamel2014, levis2015, berthier2015, petrelli2020}. However, up to our knowledge, a general effective thermal picture has not yet emerged.
 
An approach that can be exploited to test at mesoscopic level different definitions of out of equilibrium temperatures is offered by the so called {\it fluctuation theorems}, i.e. universal constraints on the probability distribution of integrated observables like work, heat and entropy production evaluated along the trajectories of individual physical entities of the system of interest \cite{gallavotti1995, spohn1999, crooks1999, harris2007, sevick2008, seifert2012, dabelow2019}. An important result showing that temperatures naturally enter heat fluctuation theorems is provided by \cite{bodineau2007}, in which it was shown that the heat exchanged between two equilibrium thermal baths satisfies the following fluctuation theorem
\begin{equation}
I(-q)-I(q)=\left(\frac{1}{T_1}-\frac{1}{T_2}\right)q~,
\label{eq:ft_original}
\end{equation} 
where $q$ is the heat exchanged per unit time, $I(q)\equiv\lim_{t\uparrow\infty}-\log P(q)/t$ is its associated Rate Function from Large Deviation Theory \cite{dembo1984, denhollander2000, touchette2009} and $T_1$,$T_2$ coincide with the bath temperatures. The above result was later studied in the context of Brownian particles \cite{visco2006, fogedby2011}, finding that its validity is in general restricted to finite intervals of $q$. The fluctuation theorem \autoref{eq:ft_original} represents a natural starting point for an investigation on the values $T_1$ and $T_2$ could take in (possibly still valid) fluctuation theorems in out of equilibrium contexts, where fluctuations are central. The temperatures defined from a fit of \autoref{eq:ft_original} and denoted as $T_{FT}$ can in fact be compared with other significant definitions of temperatures, such as the {\it effective temperature}, denoted as $T_{eff}$ and defined from the deviation of the fluctuation-dissipation theorem \cite{Ilg2006, cugliandolo2011, loi2011, isaac2011, dieterich2015, levis2015, nandi2018, cugliandolo2019}, or the {\it kinetic temperature}, instead denoted as $T_{kin}$ and defined from the equipartition theorem \cite{Ilg2006, nandi2018, mandal2019, petrelli2020} (see \autoref{app:eff_kin_temp} for more details).

Here, we numerically investigate the definition of $T_{FT}$ by considering an idealized setup which consists in a single one-dimensional particle moving in a quartic double well potential (see \autoref{fig:fig1} for a schematic depiction). In each well the particle is put in contact with a different overall thermal bath, thus experiencing a different temperature. In the right well we place an equilibrium thermal bath which is formalized through a Gaussian zero-mean delta-correlated white noise plus a viscous friction force and satisfies a usual fluctuation-dissipation theorem with an effective temperature trivially coinciding with both the bath and the kinetic ones. For the left well bath we instead consider two different cases: first we fix a further equilibrium thermal bath with same characteristics as the one in the right well expect for a different temperature; then we make it an active bath by fixing an equilibrium thermal bath analogous to the one from the right well, now with same temperature, and introducing an Ornstein-Uhlenbeck process playing the role of an additional Gaussian coloured noise with exponential self-correlation. In this way, when in the left well, the particle effectively turns into an active particle, more specifically an Active Ornstein-Uhlenbeck particle \cite{bonilla2019, caprini2021, cates2021, semeraro2021}. Moreover we remark that in this case neither the usual fluctuation-dissipation theorem is satisfied \cite{caprini2021b} nor kinetic and effective temperatures coincide \cite{petrelli2020}. The heat exchanged is measured according to two different definitions: as the work performed by one of the two baths on the particle and as the sum of the kinetic energies carried by the particle every time it jumps from one well to the other. The first definition is nothing but the usual heat as defined in the framework of stochastic thermodynamics \cite{sekimoto2010, peliti2021}. The second one is instead newly introduced as suggested by our specific setup.

We find that in all cases considered the fluctuation theorem \autoref{eq:ft_original} is still valid. More in detail, in the case in which two equilibrium thermal baths with different temperatures are fixed, both definitions of heat exchanged lead to the validity of \autoref{eq:ft_original} with slope in accordance with the bath temperatures. This first result provides an essential correspondence between $T_{FT}$ and both $T_{eff}$ and $T_{kin}$ which, as mentioned above, in this case both trivially coincide with the bath ones. In the active bath case, we find instead that different values of $T_{FT}$ associated to the active bath emerge based on the definition of heat under study. When the heat as sum of kinetic energies is considered, the extracted $T_{FT}$ turns out to correspond to the kinetic temperature of the active bath. When instead considering the heat as the work performed by the thermal environment, this temperature assumes intermediate values between the kinetic temperature and the effective one. These results and discrepancies can be interpreted by looking at the particle jump phenomenology.

The remainder of the paper is structured as follows. In \autoref{sec:model_general} we present the model and methods we adopted. In particular, in \autoref{sec:model} we describe our setup and detail the two cases under scrutiny, in \autoref{sec:heat_defs} we introduce the two definitions of heat exchanged we consider along with the energy balance of the system, in \autoref{sec:num_methods} we describe the numerical methods we adopted and in \autoref{sec:stat_pos} we comment on the stationary position distribution of the system. Next, in \autoref{sec:results_fr_temps} we present and comment the results of our investigation for the two bath configurations considered. Finally, in \autoref{sec:conclusions} we report our closing remarks.

\section{Model and Methods}
\label{sec:model_general}

\subsection{Model}
\label{sec:model}

\begin{figure}[t]
\begin{center}
  \begin{tabular}{cc}
       \includegraphics[width=0.65\columnwidth]{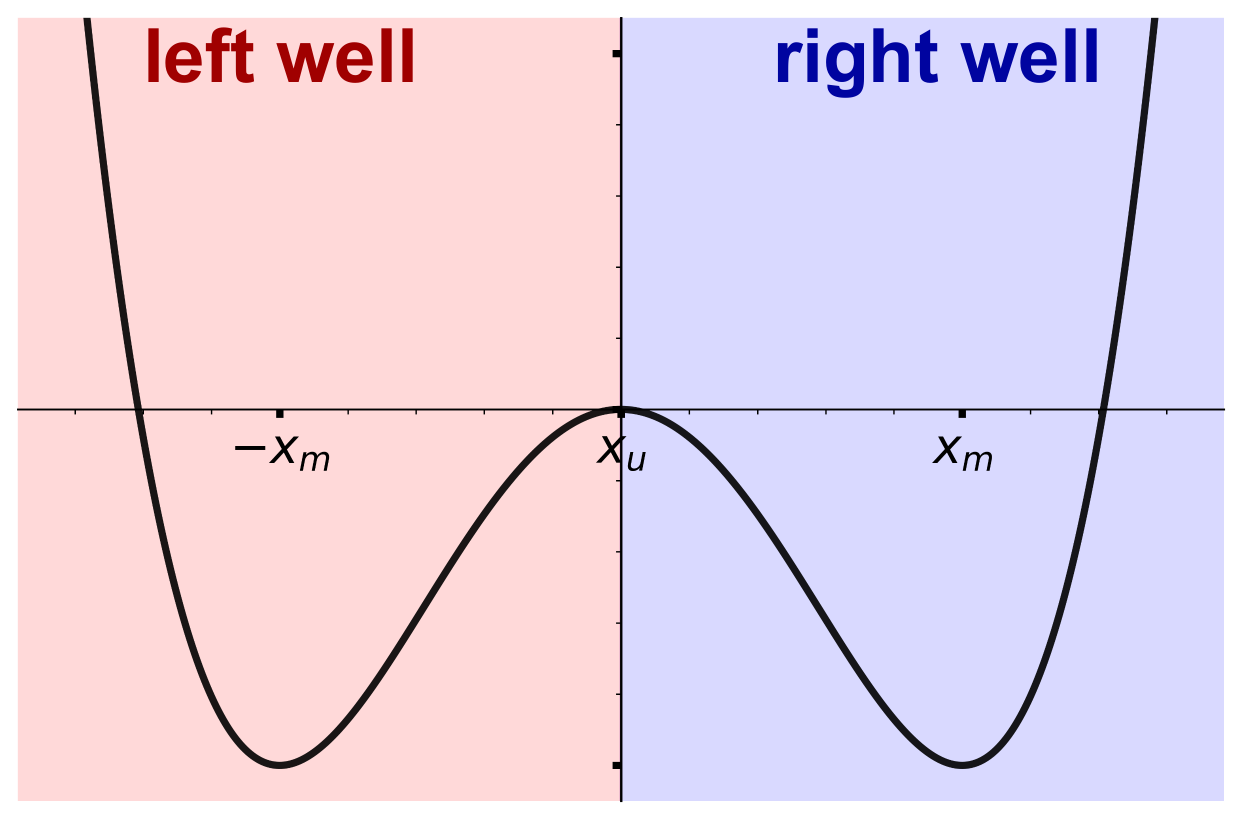}
  \end{tabular}
\caption{\footnotesize{Schematic depiction of our idealized setup. The black line denotes the quartic double well potential \autoref{eq:potential} with minima and local maxima at $\pm x_m$ and $x_u$, respectively, and depth $\Delta U$, the red and blue areas and labels below and above $x_u$ denote the action of baths with different features in the two wells and the gray circle denotes the Brownian particle while jumping from the left to the right well, as suggested by the black arrow.}}
\label{fig:fig1}
\end{center}
\end{figure}

The general framework of our setup is that of a unidimensional unit-mass mesoscopic particle of position $x(t)$ and diameter $\sigma=1$ moving under the action of the external quartic double well potential
\begin{equation}
U(x(t))=\frac{a}{4}~(x(t)-x_u)^4-\frac{b}{2}(x(t)-x_u)^2~,
\label{eq:potential}
\end{equation}
where $a,b>0$ and $x_u$ is the centre of the potential which we set to zero and serves as separating point between the regions $x> x_u$ ({\it right well}) and $x\leq x_u$ ({\it left well}) where thermal baths with different features act. The local maximum of the potential is located at $x_u=0$, the global minima are at $\pm x_m=x_u\pm\sqrt{b/a}$ distanced by $2\sqrt{b/a}$ and its depth $\Delta U=U(x_u)-U(\pm x_m)=b^2/4a$ represents the height of the barrier the particle has to overcome to hop from one well to the other. In order to highlight the spatial separation of the two baths induced by the potential \autoref{eq:potential}, we recast the usual Langevin equation describing the particle dynamics with initial conditions $x(0)\equiv x_0$ and $\dot{x}(0)\equiv v_0$ into the following form
\begin{equation}
\ddot{x}(t)=B_1(\dot{x}(t), t)\theta(x(t))+B_2(\dot{x}(t),t)(1-\theta(x(t)))-\frac{dU[x(t)]}{dx}~,
\label{eq:general_langevin}
\end{equation}
 where $B_{1}(\dot{x}(t),t)$ and $B_{2}(\dot{x}(t),t)$ collect the forces exerted by the overall baths in the two wells. The presence of the Heaviside functions $\theta(x(t))$ ensures in fact $B_1(\dot{x}(t),t)$ and $B_2(\dot{x}(t),t)$ to act only when the particle is in the right or left well, respectively\footnote{For the sake of simplicity, here we assume the convention $\theta(0)=0$ \cite{kanwal2012} instead of the half-maximum one $\theta(0)=1/2$ \cite{abramowitz1972} so that the function is left-continuous at $x=0$ and $\theta(t)$ and $1-\theta(t)$ can be properly considered as the indicator functions of the intervals $(0,+\infty)$ and $(-\infty,0]$, respectively. We underline that this choice does not affect our results as the value of a function at a single point does not affect the overall values of the heat integrals from \autoref{sec:heat_defs}.}. Concerning instead the action of the baths, whenever the particle hops into each of the two wells, their corresponding noise processes are made to restart acting with an initial condition extracted from their stationary distributions. \autoref{fig:fig1} graphically summarizes our setup highlighting with different colours the left and right well regions.

We now specify the actual composition of the forces contributing to each bath. In the right well, $B_1(\dot{x}(t),t)$ is always associated to a usual equilibrium thermal bath, hereafter referred to as {\it passive bath}, thus
\begin{equation}
B_1(\dot{x}(t),t)=-\gamma \dot{x}(t)+\sqrt{2\gamma T_1}~\xi_1(t)~,
\label{eq:b1}
\end{equation} 
where $\gamma$ is the viscous friction coefficient, $T_1$ is the bath tempexrature and $\xi_1(t)$ is a usual Gaussian white noise with $\braket{\xi_1(t)}=0$ and $\braket{\xi_1(t)\xi_1(t')}=\delta(t-t')$. Note that, for the sake of simplicity, here and in the following we set the Boltzmann constant $k_B$ to unity. The distribution for the restart of $\xi_1(t)$ is then a normal Gaussian $\mathcal N(0,1)$. As aforementioned, for the bath in the left well we instead distinguish two different cases:
\begin{enumerate}[label=\alph*)]
\item another passive bath with friction coefficient $\gamma$ and temperature $T_2$, i.e.
\begin{equation}
B_2(\dot{x}(t),t)=-\gamma \dot{x}(t)+\sqrt{2\gamma T_2}~\xi_2(t)~,
\end{equation} 
where $\xi_2(t)$ is a Gaussian white noise independent on $\xi_1(t)$ with $\braket{\xi_2(t)}=0$, $\braket{\xi_2(t)\xi_2(t')}=\delta(t-t')$ and $T_2$ in general different from $T_1$. As for $\xi_1(t)$, the distribution for the restart of $\xi_2(t)$ is $\mathcal N(0,1)$. Note that in this specific case the temperature for the entire domain can be written as the $x$-dependent function $T(x)\equiv T_2+(T_1-T_2)\theta(x)$, so that the overall Langevin equation \autoref{eq:general_langevin} can be recast as
\begin{equation}
    \ddot{x}(t)=-\gamma \dot{x}(t)+\sqrt{2\gamma T[x(t)]}~\xi(t)-\frac{dU[x(t)]}{dx(t)}~,
    \label{eq:langevin_mult}
\end{equation}
where $\xi(t)$ is a single Gaussian white noise with $\braket{\xi(t)}=0$ and $\braket{\xi(t)\xi(t')}=\delta(t-t')$ acting everywhere in the system which is made multiplicative by the presence of $T(x)$ in its multiplicative factor;

\item a passive bath with friction coefficient $\gamma$ and temperature $T_2$ and an additional {\it Ornstein-Uhlenbeck noise} reminiscent of the active force from the active Ornstein-Uhlenbeck particle model \cite{bonilla2019, caprini2021, cates2021, semeraro2021} and hereafter referred to as {\it active bath}, i.e. 
\begin{equation}
B_2(\dot{x},t)=-\gamma \dot{x}(t)+\sqrt{2 \gamma T_2}~\xi_2(t)+a(t)~,
\end{equation} 
 where $\xi_2(t)$ is a Gaussian white noise analogous to the one from case a) and $a(t)$ is an Ornstein-Uhlenbeck process implemented as the solution of the additional stochastic differential equation
\begin{equation}
\dot{a} (t)=-\gamma_R a(t)+F_a\sqrt{2 \gamma_R}~\eta(t)
\label{eq:orn_uhl}
\end{equation} 
with initial condition $a(0)\equiv a_0$, where $\eta(t)$ is a further Gaussian white noise independent on both $\xi_1(t)$ and $\xi_2(t)$ with $\braket{\eta(t)}=0$ and $\braket{\eta(t)\eta(t')}=\delta(t-t')$ and $\gamma_R^{-1}$ and $F_a$ are the {\it persistence time} associated to the the active process and a positive constant ruling its magnitude, respectively. From the average and self-correlation of $a(t)$
\begin{equation}
\braket{a(t)}=a_0e^{-\gamma_R t}\qquad\text{and}\qquad\braket{a(t)a(t')}=a_0^2e^{-\gamma_R(t+t')}+F_a^2\left(e^{-\gamma_R|t-t'|}-e^{-\gamma_R(t+t')}\right)~,
\label{eq:av_corr_a}
\end{equation}
one in fact immediately realizes that $\tau_p=\gamma_R^{-1}$ controls the exponential decay of both average and self-correlations and that at large times $\braket{a^2(t)}\simeq F_a^2$, so that $F_a$ indeed plays the role of average magnitude for the active process \cite{caprini2019, caprini2021, semeraro2021}. \autoref{eq:av_corr_a} also suggests that the distributions for the restart of $\xi_2(t)$ and $a(t)$ are $\mathcal N(0,1)$ and $\mathcal N(0,F_a^2)$, respectively. In order to better discern the action of $a(t)$, here we fix $T_1=T_2$ and, for the sake of simplicity we also set $\gamma_R=3T_2/(\gamma \sigma^2)$ \cite{das2018, caporusso2020, semeraro2023a}. Moreover, as typically done \cite{fily2014, cates2015, das2018, mandal2019, caporusso2020, semeraro2023a}, we control the relative magnitude activity-thermal noise varying the adimensional P\'eclet number
\begin{equation}
    Pe\equiv\frac{F_a\sigma}{T_2}~,
    \label{eq:pe_aoup}
\end{equation}
where we recall $\sigma=1$ to be the particle diameter. We remark that in general, the active bath configuration can be realized in actual experiments making use of Janus particles \cite{jiang2010, theurkauff2012, buttinoni2013, walther2008} or and optical tweezers \cite{wang2002, blickle2006}, or also introducing a passive tracer particle in a suspension of active particles whose collisions with the tracer itself can be described by $a(t)$ \cite{isaac2015, dabelow2019}.

\end{enumerate}

Finally, in order to allow the particle to correctly thermalize in each well before every jump, we need to correctly assess the relevant timescales of the system. Concerning case a), there are only two relevant timescales. The first one is the {\it inertial time} $\tau_I=\gamma^{-1}$, which is the typical time needed to attain thermal equilibrium with the bath. The second one is the average time the particle remains in one well before hopping starting the barrier ascension from $\pm x_m$, or {\it average residence time}, $\tau_r$. In the overdamped limit for a single white-noise bath acting everywhere and a parameter choice such that $\Delta U/T\gg 1$, $\tau_r$ is estimated as \cite{hanggi1990}

\begin{equation}
\tau_r=\frac{\pi\gamma}{\sqrt{U''(x_m)|U''(x_0)|}}e^{\frac{\Delta U}{T}}=\frac{\pi\gamma}{\sqrt{2}b}e^{\frac{\Delta U}{T}}~,
\label{eq:resid_time}
\end{equation} 

where $U''(x(t))$ is the second derivative of the potential \autoref{eq:potential}. In order to allow the particle to thermalize after each jump we require $\tau_r > \tau_I$ in each well. In the following we use the symbols $\tau_r^l$ and $\tau_r^r$ to denote the average residence times in the left and right well, respectively. Concerning case b) yet another timescale needs to be considered: the persistence time $\tau_p=\gamma_R^{-1}$ controlling the exponential decay of the coloured noise correlations. A further condition that is required to let the particle thermalise in presence of the additional Ornstein-Uhlenbeck force is then $\tau_r>\tau_p$. We remark that in presence of an active process like $a(t)$ a Kramers-like formula similar to \autoref{eq:resid_time} for $\tau_p$ is still in place in some limiting conditions \cite{caprini2021c}. However, we checked in our settings that such a formula does not hold, thus forcing us to resort to numerical estimations.

\subsection{Definitions of Heat Exchanged and Energy Balance}
\label{sec:heat_defs}

Our primary interest focuses on the heat exchanged between the two wells as the particle hops between them, which here we sample according to two different definitions capturing each different physical aspects of the system.

The first definition we consider relies on the intuitive idea that exchanges of energy and heat between the two baths must be somehow be related to the jumps of the particle from one well to the other. More in detail, each of the $N_E\geq0$ jumps occurring during a time interval of duration $\tau$ can be considered as an event of instantaneous transfer of kinetic energy from one bath to the other, with the particle playing the role of carrier. Therefore, an intuitive way in which we define the energy exchange between the two is as
\begin{equation}
\Q_E^R\equiv\frac{1}{2}\sum_{j=1}^{N_E}\big|\dot{x}(\tau_j)\big|\dot{x}(\tau_j)~.
\label{eq:qe_def}
\end{equation}
We would like to stress that the above formula is simply configured as the sum of the kinetic energies carried during each jump by the particle, i.e. a simple quantitative version of the intuitive idea delineated above. Here the subscript $E$ denotes the energetic origin of this definition, while $\{\tau_j\}_{j=1,\ldots,N_E}$ is the succession of times during the sampling interval of duration $\tau$ in which all jumps events occur, i.e. at which $x(t)=x_u$. Note that the absolute value in \autoref{eq:qe_def} ensures the increments of $\Q_E$ to be given a proper sign depending on the direction of each jump event. For the right well they are in fact positive (negative) when the particle jumps from left (right) to right (left), coherently with the physical intuition that the right well bath receives (loses) energy when the particles enters in (goes away from) it. In order to remain faithful to the prescription that a bath acquires (loses) energy when the particle jumps in (away from) it, when focusing on the left well we need to invert our point of view. In particular, now the increments of $\Q_E^L$ must be considered negative (positive) when the particle jumps from left (right) to right (left). In terms of the total energy exchange $\Q_E^L$, this translates into an overall minus sign with respect to $\Q_E^R$, i.e. $\mathcal{Q}_E^L=-\mathcal{Q}_E^R$. Trivially, $\mathcal{Q}^L_E+\mathcal{Q}^R_E=0$.

The second definition we consider takes up the usual one provided by stochastic thermodynamics in which heat is defined as the work performed on the particle by the passive bath, i.e. viscous friction force plus white noise \cite{sekimoto2010, peliti2021}. In our specific setting, the definition for the heat exchanged between particle and passive bath during a time interval of duration $\tau$ in the right well transforms into 
\begin{equation}
\begin{split}
    \Q_W^R&\equiv -\int_0^\tau~B_1(\dot{x}(s),s)\theta(x(s))~\circ dx(s)\\
    &=-\sum_{j=1}^{N_R}\int_{\tau_{0,j}}^{\tau_{R,j}}~B_1(\dot{x}(s),s)\dot{x}(s)~ds \\
          &=-\sum_{j=1}^{N_R}\int_{\tau_{0,j}}^{\tau_{R,j}}~(-\gamma \dot{x}(s)+\sqrt{2\gamma T_1}~\xi_1(s))\dot{x}(s)~ds
\end{split}
\label{eq:qw_def}
\end{equation}
where the symbol $\circ$ denotes the Stratonovich prescription is adopted \cite{stratonovich1967}, the sign minus denotes that it is the particle to perform work on the bath and ensures the same sign convention as for $\mathcal{Q}^R_E$ to be fulfilled, the subscript $W$ highlights the thermodynamical origin of this definition, $N_R$ denotes the number of times the particle resides in the right well and $\tau_{0,j}, \tau_{R,j},~{j=1,\ldots,N_R}$ denote the beginning and ending times of the $j$-th residency in the well, respectively, with $\tau_{R,j}-\tau_{0,j}>0$ its duration. 

\begin{figure}[t]
\begin{center}
  \begin{tabular}{cc}
       \includegraphics[width=0.99\columnwidth]{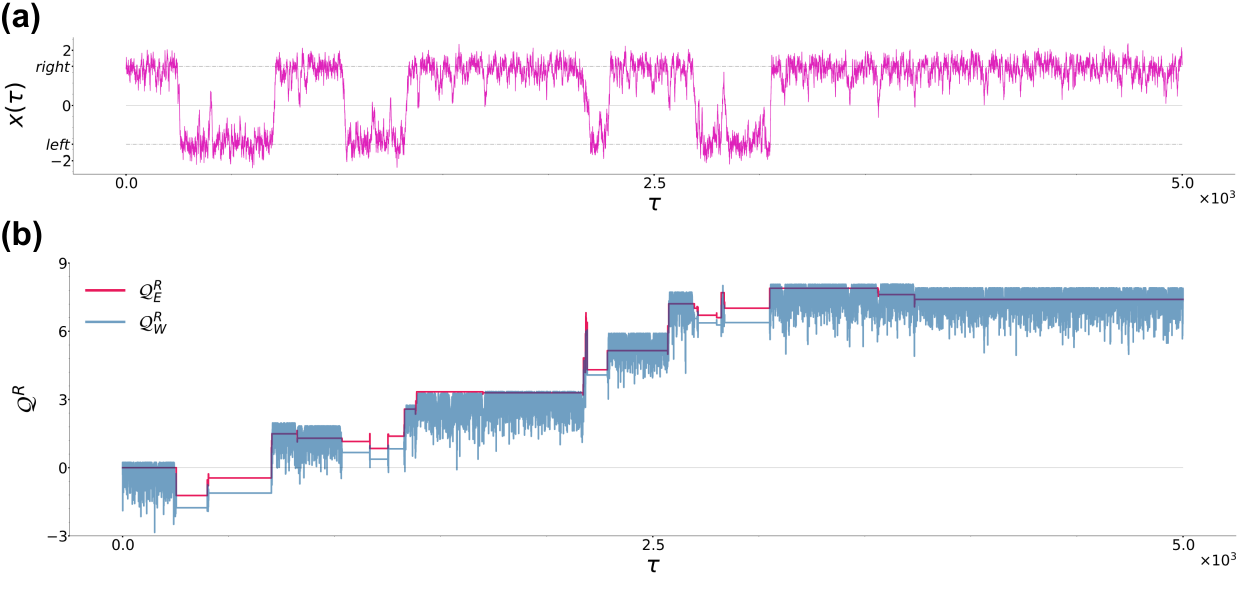}
  \end{tabular}
\caption{\footnotesize{$\textbf{(a):}$ typical trajectory of a Brownian particle from case a) at sampling time $\tau=5\cdot10^3$. The black dashed lines denote the location of the left and right potential minima at $\pm x_m=\pm\sqrt{b/a}=\pm\sqrt{2}$. $\textbf{(b):}$ time evolution of $\Q_E^R$ and $\Q_W^R$ corresponding to the trajectory in panel $\textbf{(a)}$. Parameters are $a=1.0$, $b=2.0$, $\gamma=10$, $T_1=0.2$ and $T_2=0.3$. }}
\label{fig:model_pos_distr}
\end{center}
\end{figure}

In order to provide physical intuition about the difference between the heat definitions from \autoref{eq:qe_def} and \autoref{eq:qw_def}, in panel $\textbf{(a)}$ of \autoref{fig:model_pos_distr} we show a typical particle trajectory in case a), while in panel $\textbf{(b)}$ we show the corresponding realisations of $\Q_E^R$ and $\Q_W^R$ during the same time interval (numerical data are obtained using the numerical techniques described in \autoref{sec:num_methods}). Note that $\Q_E^R$ is piecewise continuous and presents discontinuous variations only when jump events occur, while $\Q_W^R$ continuously evolves when the particle is in the right well remaining constant when the particle jumps in the left well and showing significant variations only when jump events occur. Note also that during the first permanence of the particle in the left well $\mathcal Q_W^R$ averages to zero, coherently with the fact that during this time interval the particle is thermalized with the right-well bath and the latter has not yet received any energy injection from the left-well bath. 

Following standard procedures, from \autoref{eq:qw_def} the trajectory-wise energy balance of the system can be obtained \cite{sekimoto2010, peliti2021}. By simply using the Langevin equation \autoref{eq:general_langevin} to replace $B_1(\dot{x}(t),t)\theta(x(t))$ in \autoref{eq:qw_def} for generality in case b) and adopting the Stratonovich prescription to perform integrals \cite{stratonovich1967}, one in fact finds
\begin{equation}
    \begin{split}
    \frac{1}{2}\Delta \dot{x}^2(\tau)+\Delta U(x(\tau))&=\int_0^\tau {B_1(\dot{x}(s),s)\theta(x(s))\dot{x}(s)}~ds+\int_0^\tau {B_2(\dot{x}(s),s)(1-\theta(x(s))\dot{x}(s)}~ds\\
    &=-\Q_W^R-\Q_W^L+\mathcal{W}_a~,
    \end{split}
    \label{eq:en_bal}
\end{equation}
where $\int_0^\tau~\ddot{x}(s)\dot{x}(s)~ds=(\dot{x}^2(\tau)-\dot{x}^2(0))/2\equiv\frac{1}{2}\Delta \dot{x}^2(\tau)$ and  $\int_0^\tau~\frac{dU(x(s))}{dx(s)}\dot{x}(s)~ds=(U(x(\tau))-U(x(\tau)))\equiv\Delta U(x(\tau))$ respectively denote the variation of kinetic and potential energy from initial configuration at $t=0$ and final one at $t=\tau$, $\Q_R^W$ denotes the work performed by the particle on the right passive bath defined in \autoref{eq:qw_def}, and
\begin{equation}
\begin{split}
    \Q_W^L&\equiv -\int_0^\tau~(-\gamma\dot{x}(s)+\sqrt{2\gamma T_2}~\xi_2(s))(1-\theta(x(t)))~\circ dx(s)\\
          &=-\sum_{j=1}^{N_L}\int_{\tau_{0,j}}^{\tau_{L,j}}~(-\gamma \dot{x}(s)+\sqrt{2\gamma T_2}~\xi_2(s))\dot{x}(s)~ds
\end{split}
\label{eq:qw_def_left}
\end{equation}
denotes the work performed by the particle on the passive component of the left bath only, i.e. friction force plus white noise, with $N_L$ the number of times the particle resides in the left well and $\tau_{0,j}<\tau_{L,j}$ beginning and ending times of each of the $j$-th residencies, and finally
\begin{equation}
    \mathcal{W}_a\equiv \int_0^\tau~a(s)(1-\theta(x(s)))~\circ dx(s)=\sum_{j=1}^{N_L}\int_{\tau_{0,j}}^{\tau_{L,j}}~~a(s)\dot{x}(s)~ds
    \label{eq:act_work}
\end{equation}
denotes the active work, i.e. the work performed by the additional noise in the left well returning the energy cost to sustain the particle self propulsion \cite{pietzonka2019, dabelow2019, keta2021, semeraro2021, semeraro2023a}. In both \autoref{eq:qw_def_left} and \autoref{eq:act_work} $\circ$ again underlies the Stratonovich prescription. We point out that $\Q_W^{R,L}$ and $\mathcal W_a$ are energy contributions extensive in time, while the variation of both kinetic and potential energies $\Delta\dot{x}^2(\tau)/2$ and $\Delta U(x(\tau))$ are not, i.e. 
\begin{equation}
    \lim_{\tau\uparrow\infty}\frac{1}{\tau}\int_0^\tau d\left(\frac{1}{2}\dot x^2(t)+U(x(t))\right)=0~,
    \label{eq:non_ext}
\end{equation}
or, by assuming ergodicity,
\begin{equation}
    \frac{d}{dt}\Braket{\frac{1}{2}\dot x^2(t)+U(x(t))}=0~,
    \label{eq:non_ext}
\end{equation}
where the derivative is zero due to $\braket{\dot x^2(t)/2+U(x(t))}$ assuming a constant value independent of $t$.
As a consequence, when passing to the energy balance per unit time, at times much larger than all relevant timescales of the system one has
\begin{equation}
    0=-q_W^R-q_W^L+w_a~.
    \label{eq:en_bal_qe}
\end{equation}
where $q=\Q/\tau$ for all sub- and superscripts and $w=\mathcal W_a/\tau$. Note that, coherently with the fact that the system under consideration is globally isolated, \autoref{eq:en_bal_qe} shows the overall energies exchanged by the two baths, $-q_W^R$ for the right and $-q_W^L+w_a$ for the left one, to be of opposite sign and sum to zero.

Finally, a few comments and remarks. We underline that case a) does not include the additional noise $a(t)$, so that $\mathcal W_a=0$ and \autoref{eq:en_bal} reduces to the usual equivalence between energy variation of the system and heat exchanged. We would also like to stress that in case b) $\Q_W^L$ does not capture the heat exchanges related to the left bath in its entirety. As -$\int_0^\tau B_2(\dot{x}(s),s)\dot{x}(s)(1-\theta (x(s)))~ds=\Q_W^L-\mathcal W_a$, the latter in fact also includes the active work contribution. Nevertheless, as shown in \autoref{sec:case_b}, $\Q_W^L$ is indirectly influenced by the action of the active noise as the latter clearly affects the particle velocity in the left well. In this respect, we remark that, as the active noise $a(t)$ pushes the particle, it is very likely for $a(t)$ and $\dot{x}(t)$ to have the same sign so that $\mathcal{W}_a$ from \autoref{eq:act_work} is very unlikely to be negative. Finally, referring to the trajectory of $\mathcal Q_W^R$ relative to case a) from \autoref{fig:model_pos_distr}$\textbf{(b)}$, we conclude by pointing out that during each permanence of the particle in the right well $\mathcal Q_W^R$ is bounded from above by a different value. In order to prove this point, let us consider a particle which has jumped into the right well the last time at $\tau_{J}$ and up to time $\tau>\tau_J$ remained in it. Then one has
\begin{equation}
        \begin{split}
        \mathcal Q_W^R&=c_q-\int_{\tau_J}^{\tau}~(-\gamma \dot{x}(s)+\sqrt{2\gamma T_1}~\xi_1(s))\dot{x}(s)~ds\\
                      &=\left(\frac{1}{2}\dot{x}^2(\tau_J)+U(x(\tau_J))\right)-\left(\frac{1}{2}\dot{x}^2(\tau)+U(x(\tau))\right)+c_q\\
                      &\leq \frac{1}{2}\dot{x}^2(\tau_J)+U(x(\tau_J))-U_m+c_q~,
        \label{eq:bound}
        \end{split}
\end{equation}
where in the first row we used the definition \autoref{eq:qw_def} and $c_q$ records the value accumulated by $\mathcal Q_W^R$ up to time $\tau_J$, in the second row we instead used the Langevin equation \autoref{eq:general_langevin} and performed integration similarly to the case of \autoref{eq:en_bal} and finally in the third row we used the lower bounds $\dot{x}^2(\tau)/2\geq 0$ and $U(x(\tau))\geq U(\pm x_m)\equiv U_m$. Interestingly, the bound of $\mathcal Q_W^R$ from \autoref{eq:bound} comes to depend depend on $c_q$, which englobes the integration of $\mathcal Q_W^R$ up to time $\tau_J$, on the potential energy $U_m$ at the minimum and also on the kinetic and potential energies evaluated exactly at $\tau_J$, i.e. when the particle last entered into the right well.

\subsection{Numerical Methods and Parameters}
\label{sec:num_methods}

The numerical integration of \autoref{eq:general_langevin} and \autoref{eq:orn_uhl} is performed via the velocity-Verlet \cite{tuckerman2023} and the Euler-Maruyama \cite{kloeden1992} integrators, respectively, with integration timestep $dt=10^{-2}$ in both cases. We choose for the quartic potential \autoref{eq:potential} $a=1$ and $b=2$, setting then a distance between minima and a barrier height of $2\sqrt{2}$ and $\Delta U=1$, respectively. Along with the unitary particle mass and diameter $\sigma$, the barrier height $\Delta U$ set the reduced units of our simulations. In all cases we fix $\gamma=10$ and $T_1=0.2$, while in case b) we fix $\gamma_R=3T_2/(\gamma\sigma^2)$ with $T_2=T_1=0.2$ and vary $Pe$ by acting on $F_a$. The inertial, persistence and right residence times therefore are $\tau_I=0.1$, $\tau_p\sim 16.67$ and $\tau_r^r=1.65\cdot 10^3$. The specific choices for $T_2$ in case a) and $F_a$ in case b) and consequent left well residence times will be instead specified case by case. We evolve the system for time intervals of duration $\tau$, in the following referred to as sampling time, up to $\tau=3\cdot 10^4$, the latter in all cases considered resulting much larger than all of the characteristic times of the system.

We sampled the heat per unit time $q=\mathcal Q/\tau$ for different sampling times $\tau$ according to the two definitions \autoref{eq:qe_def} and \autoref{eq:qw_def} (as in the definitions from \autoref{sec:heat_defs}, in the following the subscripts $E,W$ and superscrpits $L,R$ will specify case by case which heat in which well is being considered). The heat distributions $p(q)$ are obtained considering $N_p=10^6$ independent trajectories previously evolved for a time $\tau_{eq}=10^4$ much larger than all of the characteristic time so as to always start from the stationary configuration. Taking into account that whenever $q$ satisfies a large deviation principle its distributions takes the asymptotic form $p(q)\asymp e^{-\tau I(q)}$, with $\asymp$ asymptotic equivalence symbol underlying sub-exponential contributions $c(q)$ and $I(q)$ rate function \cite{dembo1984, denhollander2000, touchette2009}, these distributions are then used to check the validity of the fluctuation theorem \autoref{eq:ft_original} by evaluating the ratio
\begin{equation}
    \begin{split}
        \frac{1}{\tau}\log\left(\frac{p(q)}{p(-q)}\right)&\asymp I(-q)-I(q)=\left(\frac{1}{T_r}-\frac{1}{T_l}\right)q~,
    \end{split}
    \label{eq:fr_qe_tr_tl}
\end{equation}
where $I(-q)-I(q)$ is the rate function difference appearing in \autoref{eq:ft_original} and $T_r$ and $T_l$ denote the temperatures associated to the right and left well, respectively. Note that the symbol $\asymp$ underlies at finite times the appearance of the ratio $(c(q)-c(-q))/\tau$ which become increasingly negligible as time flows. Operatively, the estimates for $T_r$ and $T_l$, in the following denoted as $T_{FT}$, are obtained by first evaluating the ratio in the left-hand side of \autoref{eq:fr_qe_tr_tl} with our numerical distributions $p(q)$ at different $\tau$s and then performing at each of such times a linear fit of the resulting curves. Without loss of generality, in the following we consider settings in which the slope in \autoref{eq:fr_qe_tr_tl} is positive, corresponding to $T_l>T_r$. While for case a) one intuitively expects $T_r$ ($T_l$) to coincide with $T_1$ ($T_2$) (a circumstance which is indeed verified in \autoref{sec:case_a}), for case b) we have no a priori indications for the values they could take in presence of the active bath, especially for $T_l$. In order to extract a $T_{FT}$ estimate for the left well, motivated by the results for case a), we assume $T_r=T_1$, extract $T_{FT}=T_l$ from a fit of \autoref{eq:fr_qe_tr_tl} and compare it with the effective and kinetic temperatures $T_{eff}$ and $T_{kin}$, in turn numerically sampled according to their definitions from \autoref{app:eff_kin_temp}. 

To conclude, we remark that sampling both positive and negative values of heat becomes increasingly more difficult as the difference between the relevant temperatures of the two baths is made larger. Therefore in the following we implement parameter choices for which such a sampling is numerically feasible.

\subsection{Stationary Position Distribution}
\label{sec:stat_pos}

Before presenting our results, let us comment about the stationary position distribution in the two cases under consideration. These distributions, which we recall can be obtained as the solution of the Fokker-Planck equation with time derivative set to zero \cite{riskenBOOK}, provide in fact useful insights on average residence times, in turn useful for our later discussion.

\begin{figure}[t]
\begin{center}
  \begin{tabular}{cc}
       \includegraphics[width=0.99\columnwidth]{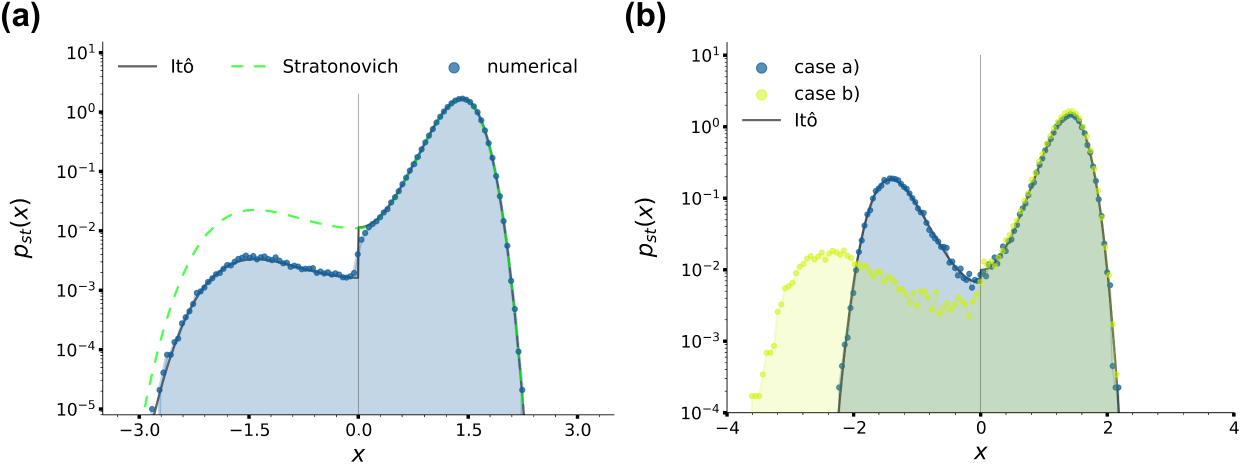}
  \end{tabular}
\caption{\footnotesize{$\textbf{(a):}$ stationary position distributions for case a) with $T_1=0.2$ and $T_2=1.4$ at sampling time $\tau=3\cdot 10^4$. The black solid and green dashed lines are the stationary solutions \autoref{eq:fpe_passive_ito} and \autoref{eq:fpe_passive_strat}, respectively, while the blue histogram is the position distribution numerically sampled, as denoted by the legend. $\textbf{(b):}$ stationary position distributions for cases a) and b) and \autoref{eq:fpe_passive_ito}, as denoted by the legend. For case a) we fix $T_1=0.2,~T_2=0.3$, while for case b) $T_1=T_2=0.2,~Pe=50$. In all panels we fixed $\gamma=10$ and $a=1.0$, $b=2.0$.}}
\label{fig:model_pos_distr_app}
\end{center}
\end{figure}

Concerning case a), in the overdamped limit and under the Itô prescription the drift and diffusion coefficients of the Fokker-Planck equation are $-\gamma^{-1}U'(x)$ and $\gamma^{-1}T(x)$ \cite{vankampen1981, riskenBOOK}, respectively, with $T(x)$ x-dependent temperature defined in \autoref{sec:model}. The resulting stationary Fokker-Planck equation has the following solution\footnote{This solution is obtained by first replacing $T(x)$ with a continuous parameter-dependent function $T_\epsilon(x)$ such that $T(x)=\lim_{\epsilon \downarrow 0}T_\epsilon(x)$, then following the standard procedure for the solution of the stationary Fokker-Planck equation with $T_\epsilon(x)$, and finally taking the limit $\epsilon\downarrow 0$.}
\begin{equation}
p_{st}(x)=\frac{N_I}{T(x)}e^{-\frac{U(x)}{T(x)}}~,
\label{eq:fpe_passive_ito}
\end{equation}
with $N_I$ normalisation factor and $U(x)$ the quartic potential \autoref{eq:potential}, which is clearly reminiscent of the equilibrium Boltzmann distribution. Note that $p_{st}(x)$ shows a jump discontinuity at $x_u$ when $T_1 \neq T_2$, which disappears when $T_1=T_2$, i.e. in the usual case of a Brownian particle under the effect of just one equilibrium thermal bath. The associated discontinuity height is $\Delta p_{st}=\big |\lim_{x\uparrow x_u}p_{st}(x)-\lim_{x\downarrow x_u}p_{st}(x)\big |=N_I\big|T_2^{-1}-T_1^{-1}\big|$ and becomes more and more marked as the difference $|T_1-T_2|$ is increased. Note also that the two temperatures $T_1$ and $T_2$ determine the shape and height of the distribution in each well, but they play no role in the maxima locations, which in turn come to coincide with the potential minima at $\pm x_m=x_u\pm\sqrt{b/a}$. For the sake of completeness, we mention that under the Stratonovich prescription the diffusion coefficient of the Fokker-Planck equation remains unaltered, while its drift coefficient becomes $\gamma^{-1}(-U'(x)+T(x))$, so that the stationary solution now is 
\begin{equation}
p_{st}(x)=N_S e^{-\frac{U(x)}{T(x)}}~,
\label{eq:fpe_passive_strat}
\end{equation}
with $N_S$ a normalisation factor, which, at variance with \autoref{eq:fpe_passive_ito}, is always continuous at $x_u$ also when $T_1\neq T_2$. We would like to stress that the difference between \autoref{eq:fpe_passive_ito} and \autoref{eq:fpe_passive_strat} can be ultimately traced back to the presence of two regions with different temperatures. In fact, as mentioned in \autoref{sec:model}, in case a) the Langevin equation \autoref{eq:general_langevin} can be recast as \autoref{eq:langevin_mult}, which is characterized by a multiplicative noise due to the presence of the $x$-dependent temperature $T(x)$. Therefore, as well known from the literature \cite{riskenBOOK, arnold2000}, applying different integration schemes leads to different results, hence the different drift coefficients for the stationary Fokker-Planck equation in the Itô and Stratonovich prescriptions and the resulting different stationary distributions \autoref{eq:fpe_passive_ito} and \autoref{eq:fpe_passive_strat}. In \autoref{fig:model_pos_distr_app}$\textbf{(a)}$ we provide a comparison between these two stationary solutions and the numerical position distributions at $\tau=3\cdot 10^4$ obtained integrating the equations of motions as described in \autoref{sec:num_methods} setting $T_1=0.2$ and $T_2=1.4$, so that $|T_1-T_2|\sim 1$. The figure at the same time shows that the numerical algorithms we use perform integration under the Itô prescription and confirms the presence of the jump discontinuity in \autoref{eq:fpe_passive_ito}.

In panel $\textbf{(b)}$ we report instead a comparison between the numerical stationary position distributions for cases a) and b). For case a) we choose $T_1=0.2$, $T_2=0.3$, while for case b) we fix $T_1=T_2=0.2$ and $F_a=10$ ($Pe=50$) so that, as will be shown in \autoref{sec:case_b}, the kinetic temperature in the left well is $\sim0.3=T_2$ (the reason for considering the kinetic temperature will appear clear in \autoref{sec:case_b}). Note that in the left well for case b) the location $-\tilde x_m$ of the peak of the distribution is shifted towards the left with respect to the location of the potential minimum $-x_m$ due to the persistent pushing of the active noise. Even though, up to our knowledge, the stationary Fokker-Planck equation in case b) has no exact analytical solution, our numerical results are coherent with the ones from \cite{caprini2019}, in which a single active Ornstein-Uhlenbeck particle in a quartic double well potential like \autoref{eq:potential} is studied. In particular, in \cite{caprini2019} the location of the peaks of the distribution are identified as the points in which the confining force due to the quartic potential \autoref{eq:potential} and the active force approximated by its average magnitude $F_a$ are balanced, i.e. as the solutions of the equation $-ax^3+bx=\pm F_a$, where the $\pm$ signs apply to the right and left wells, respectively. In our case, the solution of the above equation relative to the left well gives $-\tilde x_m\simeq 2.46$, which is in good agreement with the location of the left peak from panel $\textbf{(b)}$.

To conclude, we point out that the distributions we just commented provide qualitative insights on the average residence time $\tau_r^r$ and $\tau_r^l$ of the particle in each well, which are essential information especially for case b) in which an analytic estimate for $\tau_r^l$ is not available. It is in fact intuitive to see that in general, apart from the specific distribution features, lower temperatures associated to higher peaks in the distributions imply larger residence times, and viceversa for higher temperatures. According to panel $\textbf{(b)}$, we then intuitively expect that the average residence times in the cases under consideration rank as follows: $\tau_r^r$ is the largest, $\tau_r^l$ in case a) is intermediate and finally $\tau_r^l$ in case b) is the shortest.

\section{Results}
\label{sec:results_fr_temps}

\subsection{Heat Exchanges between Two Passive Baths}
\label{sec:case_a}

\begin{figure}[t!]
\begin{center}
  \begin{tabular}{cc}
       \includegraphics[width=0.99\columnwidth]{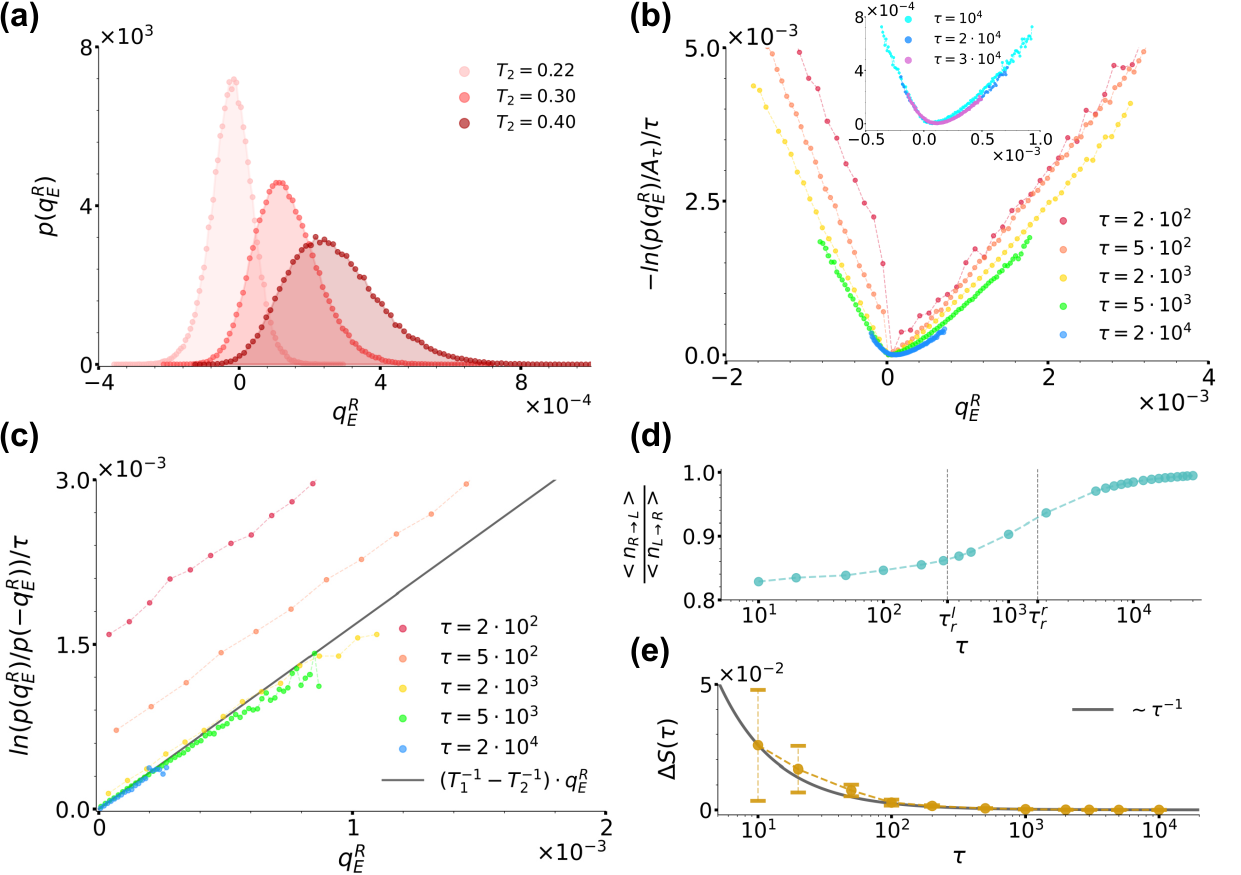}
  \end{tabular}
\caption{\footnotesize{$\textbf{(a):}$ distribution $p(q_E^R)$ for case a) at sampling time $\tau=3\cdot 10^4$ for $T_2=0.22,~0.3$ and $0.4$, as denoted by the legend. $\textbf{(b):}$ curves $-\ln(p(q_E^R)/A_\tau)/\tau$ for $T_2=0.3$ at different sampling times, as denoted by the legend. $A_\tau$ denotes the maximum of the distribution at each sampling time. The inset shows instead the trend of the same curves at the largest sampling times considered. $\textbf{(c):}$ ratio $\ln(p(q_E^R)/p(-q_E^R))/\tau$ evaluated at different sampling times $\tau$ using data from panel $\textbf{(b)}$ along with the right hand-side of \autoref{eq:fr_qe_tr_tl} plotted with $T_r=T_1=0.2$ and $T_l=T_2=0.3$, as denoted by the legend. $\textbf{(d):}$ ratio between the average number of jumps in the right$\rightarrow$left direction and left$\rightarrow$right directions denoted $\braket{n_{R\rightarrow L}}$ and $\braket{n_{L\rightarrow R}}$, respectively, as a function of sampling time. The dashed lines denote the left and right average residence times $\tau_r^l=3.11\cdot 10^2$ and $\tau_r^r=1.64\cdot 10^3$, respectively. $\textbf{(e):}$ shift $\Delta S(\tau)$ of the curves from panel $\textbf{(c)}$ as a function of time. For comparison, here the black solid line reports the trend of $\sim \tau^{-1}$. In all panels we fixed $\gamma=10$, $T_1=0.2$ and $a=1.0,~b=2.0$.}}
\label{fig:passive_baths}
\end{center}
\end{figure}

We start in this section with the investigation of case a) envisaging a passive bath in each of the two wells. We first fix $T_1=0.2$ and consider three $T_2$ values, $0.22$, $0.3$ and $0.4$. According to \autoref{eq:resid_time}, the corresponding average residence times $\tau_r^l$ are much larger than the inertial time $t_I=0.1$, ranging from $\tau_r^l=1.05\cdot 10^3$ for $T_2=0.22$ to $\tau_r^l=1.35\cdot 10^2$ for $T_2=0.4$. In \autoref{fig:passive_baths}$\textbf{(a)}$ we show the distribution $p(q_E^R)$ for these three choices of temperatures at sampling time $\tau=3\cdot 10^4$ (the distributions $p(q_E^L)$ are just symmetrical). Note that all distributions are characterised by a positive average value, $\sim 2.10~10^{-5}$ for $T_2=0.22$, $\sim 1.34~10^{-4}$ for $T_2=0.3$ and $\sim 2.68~10^{-4}$ for $T_2=0.4$, confirming that, as intuitively expected, on average the colder bath in the right well receives more energy from the hotter bath in the left well than the one it outputs towards it through the jumping particle. Note also that the distributions are characterised by an increasing skewness as the temperature difference $\Delta T=|T_1-T_2|$ is increased. In the remainder of the present section we focus on the case $T_1=0.2,~T_2=0.3$, which at the same time guarantees an appreciable skewness of $p(q_E^R)$ as well as an efficient sampling of both positive and negative heat values. Up to what our simulations afforded us to sample, we checked that the results we are about to discuss for this case also apply to the other values of $T_2$.

In order to study the validity of \autoref{eq:ft_original}, we first focus on the trend of $-\ln(p(q_E^R))/\tau$, which in the large time limit converges to the rate function $I(q_E^R)$ whenever $q_E^R$ satisfies a large deviation principle \cite{dembo1984, denhollander2000, touchette2009}. More specifically, in \autoref{fig:passive_baths}$\textbf{(b)}$ we report $-\ln(p(q_E^R)/A_\tau)/\tau$ extracted at different sampling time $\tau$s, as reported by the legend. At each sampling time, $A_\tau$ denotes the maximum of the distribution and in the ratio it makes the resulting curves shift vertically so as to have a minimum value of zero. As highlighted by the inset of \autoref{fig:passive_baths}$\textbf{(b)}$, for $\tau>10^4\gg\tau_I=0.1$ we observe that the curves do overlap, thus implying $q_E^R$ to satisfy a large deviation principle. We therefore proceed to check the validity of \autoref{eq:ft_original} as prescribed by \autoref{eq:fr_qe_tr_tl}. We use data from \autoref{fig:passive_baths}$\textbf{(b)}$ and report the resulting curves at the same $\tau$s in \autoref{fig:passive_baths}$\textbf{(c)}$, as denoted by the legend. We find that at all $\tau$s these curves are linear and, within numerical error, with slope in agreement with $1/T_1-1/T_2$, so that we can identify $T_r$ with $T_1$ and $T_l$ with $T_2$. We would like to underline that while previous results proved a fluctuation theorem like \autoref{eq:ft_original} to stand in the case of two different thermal baths separately at equilibrium but acting simultaneously everywhere in the system \cite{bodineau2007, visco2006, fogedby2011}, our results extend this scenario to the case of spatially separated baths.

Interestingly, \autoref{fig:passive_baths}$\textbf{(c)}$ also shows that at short times the numerical lines in \autoref{fig:passive_baths}$\textbf{(c)}$ do not cross the origin, but rather present a time-decreasing positive shift $\Delta S(\tau)$ which we can explain by looking at the system phenomenology at short times. During the evolution of our system, three timescales come into play, i.e. the left and right well residence times, $\tau^l_r\simeq 3.11\cdot 10^2<\tau^r_r\simeq1.64\cdot 10^3$, and the sampling time $\tau$ at which the distribution $p(q_E^R)$ is considered. When taking into account a large number$N_p$ independent realizations of the system, one then intuitively expects that, as long as $\tau<\tau_r^r$, more jumps from left to right occur than in the opposite direction, while when $\tau_r^l<\tau<\tau_r^r$, the number of right$\rightarrow$left jumps starts increasing until essentially matching the number of left$\rightarrow$right ones at $\tau\gg \tau_r^l, \tau_r^r$. \autoref{fig:passive_baths}$\textbf{(d)}$ confirms this intuition by showing the trend of the numerical ratio between the average number of jumps in the right$\rightarrow$left and in the left$\rightarrow$right directions, respectively denoted as $\braket{n_{R\rightarrow L}}$ and $\braket{n_{L\rightarrow R}}$, as a function of sampling time $\tau$. The curve in fact starts from a value lower than one for $\tau<\tau_r^l$, which then it reaches asymptotically from below when $\tau\gg\tau_r^r$. This jumps phenomenology clearly bears consequences on the distribution $p(q_E^R)$, and then on the resulting fluctuation theorem. In fact, as apparent from \autoref{fig:passive_baths}$\textbf{(b)}$, at short times $\tau<\tau_r^l$ its left and right branches weigh differently positive and negative heat values, the left branch being further away from its large-time stationary form than the right one and mirroring the jump imbalance biased towards left$\rightarrow$right positive heat jumps. As mentioned in \autoref{sec:num_methods} when commenting \autoref{eq:fr_qe_tr_tl}, these effects are encoded in the distribution as a sub-exponential contribution $c(q_E^R)$, which is a function of $q_E^R$ scaling as $t^{\alpha}$ with $\alpha<1$ and in our case is directly related with the observed shift $\Delta S(\tau)$. As shown in \autoref{fig:passive_baths}$\textbf{(e)}$, we in fact find $\Delta S(\tau)=(c(q_E^R)-c(-q_E^R))/\tau$ to decrease as $\simeq \tau^{-1}$ corresponding to $\alpha\simeq 0$, the latter value signalling that the difference $c(q_E^R)-c(-q_E^R)$ is of order $\sim\mathcal{O}(1)$.

\begin{figure}[t!]
\begin{center}
  \begin{tabular}{cc}
       \includegraphics[width=0.99\columnwidth]{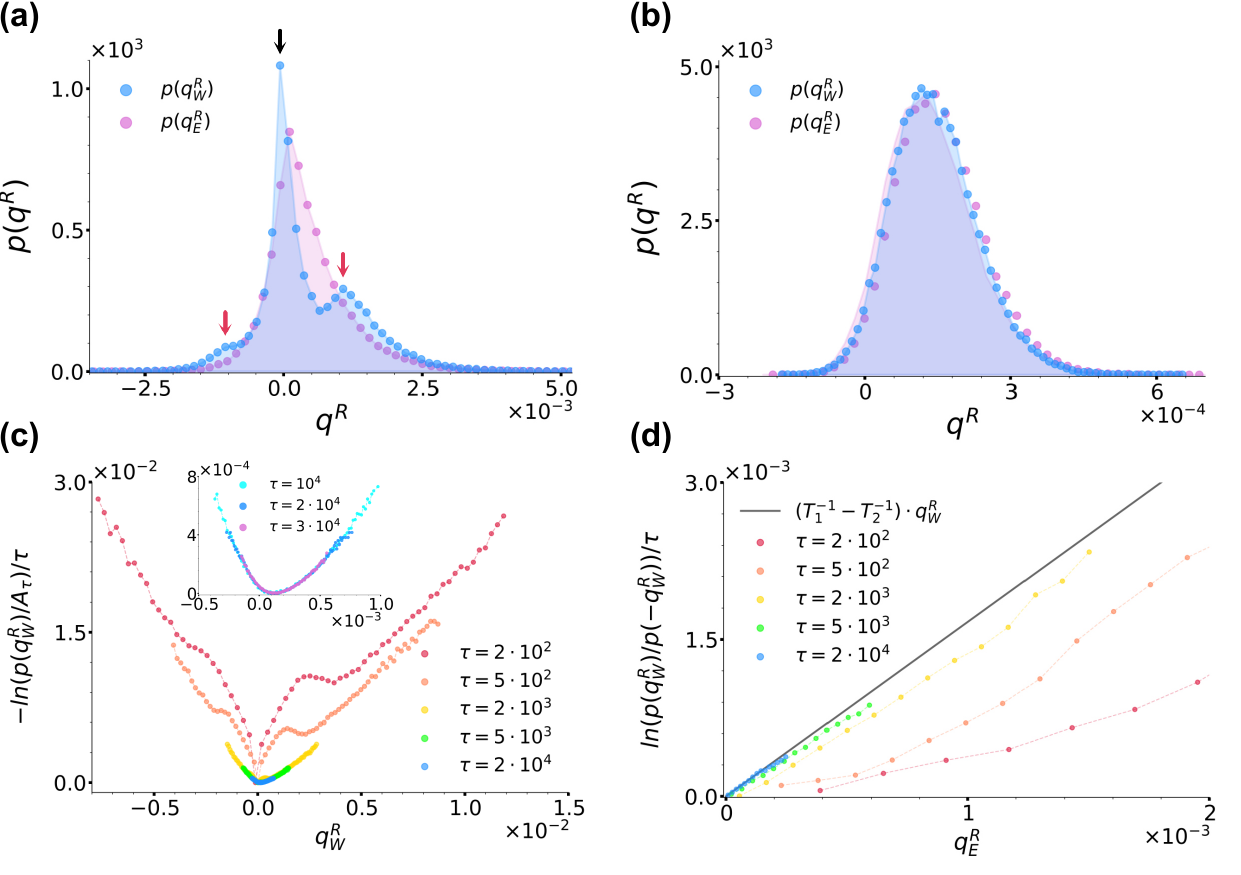}
  \end{tabular}
\caption{\footnotesize{{$\textbf{(a)}$ and $\textbf{(b):}$ comparison between the distributions $p(q_E^R)$ and $p(q_W^R)$ at sampling times $\tau=10^3$ and $\tau=3\cdot10^4$, respectively, with same parameters as in \autoref{fig:passive_baths}. In panel $\textbf{(a)}$ the three arrows highlight the three peaks of $p(q_W^R)$. $\textbf{(c):}$ curves $-\ln(p(q_W^R)/A_\tau)/\tau$ for $T_2=0.3$ at different sampling times, as denoted by the legend. As in \autoref{fig:passive_baths}, $A_\tau$ denotes the maximum of the distribution at each sampling time. The inset shows instead the trend of the same curves at the largest sampling times considered. $\textbf{(d):}$ ratio $\ln(p(q_W^R)/p(-q_W^R))/\tau$ evaluated at different sampling times using data from panel $\textbf{(c)}$ along with the right hand-side of \autoref{eq:fr_qe_tr_tl} plotted with $T_r=T_1=0.2$ and $T_l=T_2=0.3$, as denoted by the legend. In all panels we fixed $\gamma=10$, $T_1=0.2$ and $a=1.0,~b=2.0$.}}}
\label{fig:passive_baths_bis}
\end{center}
\end{figure}

We now discuss the validity of the fluctuation theorem from \autoref{eq:ft_original} for $q_W^R$ by studying $p(q_W^R)$ by comparison with $p(q_E^R)$. In \autoref{fig:passive_baths_bis}$\textbf{(a)}$ and $\textbf{(b)}$ we compare the distributions $p(q_W^R)$ and $p(q_E^R)$ for the same parameter choice as in \autoref{fig:passive_baths}$\textbf{(b)}$ at sampling times $\tau=10^3$ and $\tau=3\cdot10^4$, respectively. Let us focus on \autoref{fig:passive_baths_bis}$\textbf{(a)}$ first. What immediately catches the eye is that, at variance with $p(q_E^R)$ and as highlighted by the vertical arrows, $p(q_W^R)$ is characterized by three peaks. This peculiar structure can be readily explained by recalling the jump phenomenology discussed above. The left and right external peaks highlighted by the red arrows are due to particles leaving and entering the right well, which are then responsible for negative and positive energy exchanges, respectively. As here $\tau_r^l<\tau<\tau_r^r$, more particles have jumped from left to right than in the opposite direction, hence the higher right peak. However, in our large sample of $N_p$ independent realizations, at this time a large number of particle has not yet jumped at all from the right well, but rather have been exchanging an average zero heat with the equilibrium thermal bath in this well, hence the central peak located at $q_W^R=0$ highlighted by the black arrow. At large times, $p(q_W^R)$ instead loses its three-peaks structure and comes to coincide with $p(q_E^R)$ from \autoref{fig:passive_baths}$\textbf{(b)}$. In particular, the central peak disappears because at large times it is extremely probable that all particles have already jumped almost once, while the other two get closer and closer until eventually merging. This overall scenario is graphically confirmed and clarified by \autoref{fig:passive_baths_bis}$\textbf{(c)}$, which reports the curves $-\ln(p(q_W^R)/A_\tau)/\tau$ extracted at different sampling times $\tau$s, as denoted by the legend. The figure in fact at the same time shows the two external peaks clearly getting closer until eventually merging and also the curves converging towards a convex rate function $I(q_W^R)$. Combining further this last information with the content of \autoref{fig:passive_baths}$\textbf{(b)}$ and \autoref{fig:passive_baths_bis}$\textbf{(b)}$, we can therefore affirm that at large times $I(q_W^R)=I(q_E^R)$. The curves from \autoref{fig:passive_baths_bis}$\textbf{(c)}$ allow us to finally check the validity of a fluctuation theorem for $q_W^R$ as prescribed by \autoref{eq:fr_qe_tr_tl}. \autoref{fig:passive_baths_bis}$\textbf{(d)}$ reports the ratio $\ln(p(q_W^R)/p(-q_W^R))/\tau$ evaluated at different sampling times using data from panel $\textbf{(c)}$ and, as highlighted by the black line reporting \autoref{eq:fr_qe_tr_tl} plotted with $T_r=T_1=0.2$ and $T_l=T_2=0.3$, shows that at large sampling time $\tau$ $q_W^R$ indeed satisfies the same fluctuation theorem shown in \autoref{fig:passive_baths}$\textbf{(c)}$ to be satisfied by $q_E^R$ with same slope. Note that at short times the fluctuation theorem is not satisfied because of the sub-exponential contribution $(c(q_W^R)-c(-q_W^R))/\tau$ which encodes the three-peaks structure of $p(q_W^R)$ and makes the curve actually curvilinear rather than rigidly vertically shifted. To conclude we report that for the left well we checked that $p(q_W^L)=p(-q_W^R)$ and consequently that the same results discussed until this point for $q_W^R$ symmetrically still apply so that the energy balance \autoref{eq:en_bal_qe} results satisfied.

\subsection{Heat Exchanges between a Passive and an Active Bath}
\label{sec:case_b}

\begin{figure}[t!]
\begin{center}
  \begin{tabular}{cc}
       \includegraphics[width=0.99\columnwidth]{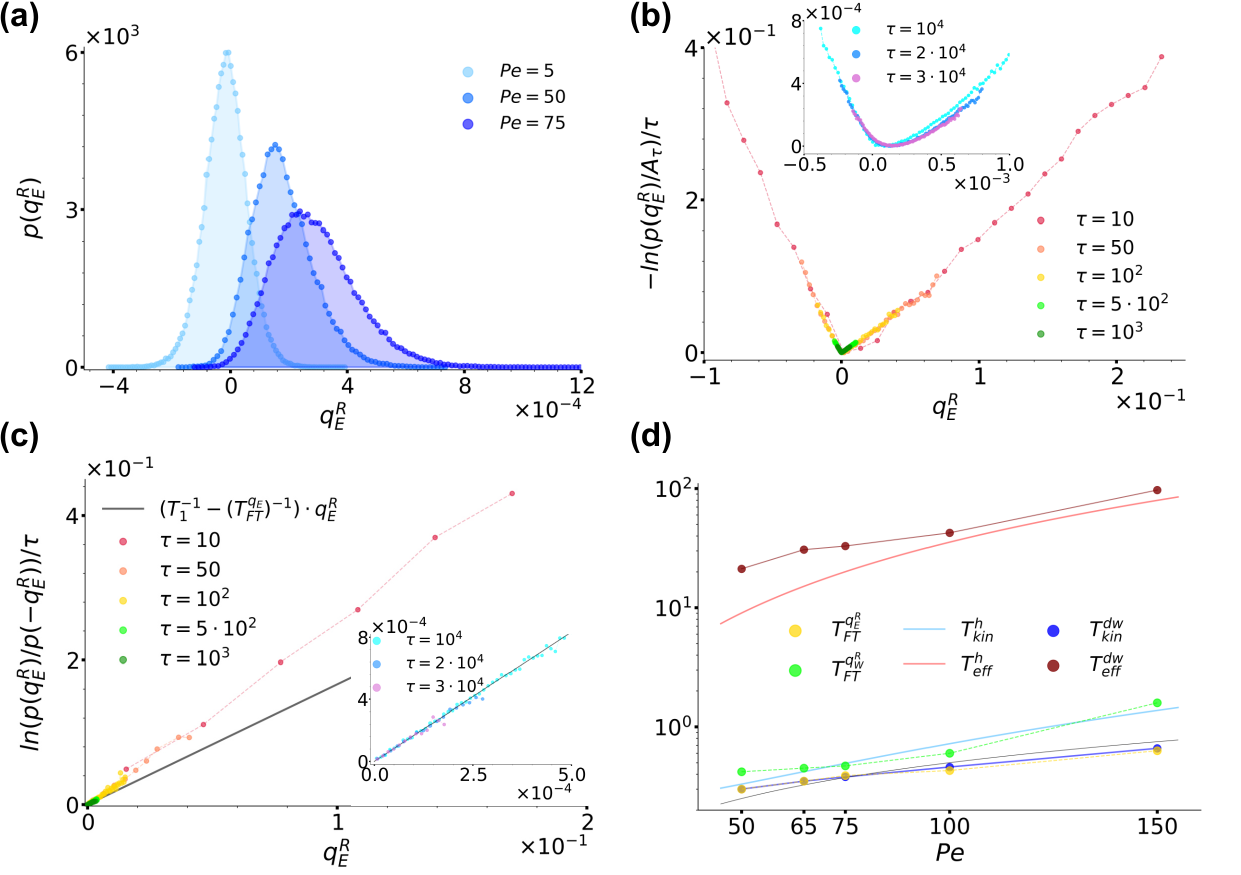}\quad
  \end{tabular}
\caption{\footnotesize{$\textbf{(a):}$ distribution $p(q_E^R)$ for case b) at sampling time $\tau=3\cdot 10^4$ for $Pe=5,~50$ and $75$, as denoted by the legend. $\textbf{(b):}$ curves $-\ln(p(q_E^R)/A_\tau)/\tau$ for $Pe=50$ at different sampling times, as denoted by the legend. As in \autoref{fig:passive_baths}, $A_\tau$ still denotes the maximum of the distribution at each sampling time. The inset shows instead the trend of the same curves at the largest sampling times considered. $\textbf{(c):}$ ratio $\ln(p(q_E^R)/p(-q_E^R))/\tau$ evaluated at different sampling times $\tau$ up to $\tau=10^3$ in the main figure and between $\tau=10^4$ and $\tau=3\cdot 10^4$ in the inset, as denoted by legends. Main and inset were obtained using data from main and inset of panel $\textbf{(b)}$, respectively, and both report the trend of $(T_1^{-1}-T_{FT}^{-1})\cdot q_E$ as a black solid line, with $T_{FT}\sim 0.3$ extracted from a fit of the curves in the inset performed as described in the main text. $\textbf{(d):}$ overview of the temperatures $T^{q_E^R}_{FT}$ and $T^{q_W^R}_{FT}$ (yellow and green circles) extracted from $p(q_E^R)$ and $p(q_W^R)$ at sampling time $\tau=3\cdot 10^4$ as prescribed by \autoref{eq:fr_qe_tr_tl} as a function of $Pe$ compared to $T_{kin}^h,T_{eff}^h$ and $T_{kin}^{dw},T_{eff}^{dw}$ obtained in the harmonic (light blue and red lines) and double-well (dark blue and red lines) configurations, respectively. $T_{kin}^h$ and $T_{eff}^h$ are plotted as solid lines to highlight their analytical origin from \autoref{eq:t_eff_h} and \autoref{eq:t_kin_h}, while all other data are plotted as dot and lines, the dots reporting the values obtained numerically, the lines being a guide to the eye, including the lower black solid line reporting a sample linear trend $\sim Pe$. In all panels we fixed $\gamma=10$ and $T_1=T_2=0.2$, while in the harmonic and double-well configurations we set $k=4.0$ and $a=1.0,~b=2.0$, respectively.}}
\label{fig:active_baths}
\end{center}
\end{figure}

In this section we investigate case b) envisaging a left bath which is given an active character through the introduction of an additional Ornstein-Uhlenbeck noise. In \autoref{fig:active_baths}$\textbf{(a)}$ we preliminarily show the distribution $p(q_E^R)$ for $T_1=T_2=0.2$ and three different $Pe$ at sampling time $\tau=3\cdot 10^4$ (the distributions $p(q_W^R)$ are just symmetrical). The figure is clearly reminiscent of \autoref{fig:passive_baths}$\textbf{(a)}$, with $Pe$ effectively playing the role of a temperature like $T_2$: as $Pe$ is increased, the distributions shift towards the right, with a consequent increase of their skewness as well as of the average value of $q_E^R$.

Let us consider in detail the case $Pe=50.0$. As in \autoref{sec:case_a}, for the right well the residence time is $\tau_r^r\sim 1.64\cdot 10^3$ as prescribed by \autoref{eq:resid_time}. For the left well we instead numerically estimate it as $\tau_r^l\sim34.18$, so that the conditions $\tau_r^{l},\tau_r^{r}>\tau_p=16.67\gg \tau_I=0.1$ are satisfied. \autoref{fig:active_baths}$\textbf{(b)}$ shows the trend of $-\ln(p(q_E^R))/\tau$ for increasing sampling time $\tau$, as denoted by the legend. As remarked by the inset and similarly to \autoref{sec:case_a}, also in this case we find the curves to converge at large times towards a convex rate function $I(q_E^R)$, thus proving $q_E^R$ to satisfy a large deviation principle even when one of the baths is made active. \autoref{fig:active_baths}$\textbf{(c)}$ shows instead the ratio $\ln(p(q_E^R)/p(-q_E^R))/\tau$ evaluated at different times using data from \autoref{fig:active_baths}$\textbf{(b)}$. Interestingly, also in this case the resulting curves show a linear trend at all times. Here the effect of the sub-exponential contribution $c(q_E^R)$ makes the slope of the curves reduce until reaching a constant value, as remarked by the inset. Following the same line of action as in \autoref{sec:case_a}, one can fit these lines as prescribed by \autoref{eq:fr_qe_tr_tl} so as to extract a temperature estimate for the right well based on fluctuation theorems. When doing so identifying a priori $T_r$ with $T_1=0.2$, one finds for the left well $T^{q_E^R}_{FT}\sim 0.3$ (whence the choice of parameters in \autoref{fig:model_pos_distr_app}$\textbf{(b)}$ for which $T_2=0.3$ from case a) and $T^{q_E^R}_{FT}\sim 0.3$ from case b) essentially coincide). The resulting line $(T_1^{-1}-(T^{q_E^R}_{FT})^{-1})\cdot q_E^R$ is reported in \autoref{fig:active_baths}$\textbf{(c)}$ for completeness. The possibility $T_l=T_2=0.2$ can thus be trivially discarded as, like in the case $Pe=5$ from \autoref{fig:active_baths}$\textbf{(a)}$ in which the effect of the active noise is essentially negligible, it would lead to $p(q_E)=p(-q_E)$, and therefore to a vanishing slope. 

At this point one could naturally ask how $T^{q_E^R}_{FT}$ is influenced by the strength of the activity, i.e. by $Pe$, and also whether this temperature does coincide with other out-of-equilibrium temperatures like the kinetic and effective ones mentioned in \autoref{sec:intro} and detailed in \autoref{app:eff_kin_temp}. Concerning the first question, \autoref{fig:active_baths}$\textbf{(d)}$ shows the trend of $T^{q_E^R}_{FT}$ (yellow dots) obtained varying $Pe$ while keeping fixed all other system parameters, while \autoref{tab:comparison_active} reports the exact values emerged from the analytical expressions from \autoref{app:eff_kin_temp} and from our fit procedure performed keeping $T_r=T_1=0.2$ fixed. The data suggest that $T^{q_E^R}_{FT}$ increases roughly linearly as a function of $Pe$. Note that at all $Pe$ considered the estimated left residence times $\tau_r^l$ satisfy the condition $\tau_r^{l}>\tau_p=16.67\gg \tau_I=0.1$, as reported in \autoref{tab:comparison_active}.

In order to answer the second question, we instead compare $T^{q_E^R}_{FT}$ with the kinetic and effective temperatures obtained for a particle subjected everywhere to the active bath and under the separate action of two different external potentials. The first potential we consider is a harmonic one $U(x)=kx^2/2$ introduced in such a way to approximate the quartic double well \autoref{eq:potential} around one if its minima, i.e setting setting $k=2b$ with $2b$ second derivative of \autoref{eq:potential} at its minima locations $\pm x_m$. In such a case we refer to the overall configuration as harmonic configuration, the two temperatures are denoted as $T_{eff}^h$ and $T_{kin}^h$ and, as detailed in \autoref{app:eff_kin_temp}, an analytical derivation is feasible with resulting expressions provided by \autoref{eq:t_eff_h} and \autoref{eq:t_kin_h}. The second potential we consider is instead the usual quartic double well potential \autoref{eq:potential}, and the relative configuration is referred to as double-well configuration. Here the two temperatures are denoted $T_{eff}^{dw}$ and $T_{kin}^{dw}$ and their estimates are obtained by numerical means. The rationale underlying these configurations follows our desire to understand if the value of $T^{q_E^R}_{FT}$ is determined mostly by the permanence of the particle around the potential minimum (hence the harmonic configuration), or, similarly to what observed for the variation of the entropy production for an active particle under the action of a quartic double-well potential \cite{caprini2019, dabelow2021}, by the non-convex region of \autoref{eq:potential} (hence the double-well configuration). 

\begin{table}[t!]
\centering
\begin{tabular}{!{\vrule width 1.5pt} c !{\vrule width 1.5pt} c !{\vrule width 1.5pt} c|c !{\vrule width 1.5pt} c|c|c|c !{\vrule width 1.5pt}}
\noalign{\hrule height 1.5 pt}
$Pe$ & $\tau_r^l$ & $T_{FT}^{q_E^R}$ & $T_{FT}^{q_W^R}$ & $T^h_{eff}$ & $T^h_{kin}$ & $T^{dw}_{eff}$ & $T^{dw}_{kin}$\\
\noalign{\hrule height 1.5 pt}
50.0  & 34.18 & 0.30  & 0.42  & 22.05   & 0.33  & 21.20  & 0.30\\
\hline
65.0  & 32.09 & 0.35  & 0.45  & 37.13   & 0.42  & 30.68  & 0.35\\
\hline
75.0  & 31.41 & 0.39  & 0.47  & 49.36   & 0.49  & 32.92  & 0.38\\
\hline
100.0 & 29.85 & 0.43  &  0.60 & 92.42   & 0.72  &  37.49 & 0.46 \\
\hline
150.0 & 27.36 & 0.63  & 1.59  & 207.69  & 1.43  & 97.12  & 0.66 \\
\hline
\noalign{\hrule height 1.5 pt}
\end{tabular}
\caption{\footnotesize{Estimates of the average residence time in the left well $\tau_r^l$ and of the temperatures $T_{q_E^R}$, $T_{q_W^R}$, $T_{eff}^h$, $T_{kin}^h$, $T_{eff}^{dw}$, $T_{kin}^{dw}$ for various choices of increasing $Pe$. The values of $T_{eff}^h$, $T_{kin}^h$ are obtained analytically from \autoref{eq:t_eff_h} and \autoref{eq:t_kin_h}, while all other time and temperature estimates are obtained numerically. In all cases the system was evolved until $\tau=3\cdot 10^4$ and we fixed $\gamma=10$ and $T_1=0.2$, $T_2=0.2$, while in the harmonic and double-well configurations we set $k=4.0$ and $a=1.0,~b=2.0$, respectively.}}
\label{tab:comparison_active}
\end{table}

\autoref{fig:active_baths}$\textbf{(d)}$ reports the trend of the temperature values we obtained in the two configurations under consideration, while \autoref{tab:comparison_active} offers an overview of their values. From a comparison with the trends and values of $T_{FT}^{q_E^R}$ it is immediate to realise that there is no correspondence between any of the two effective temperatures. The kinetic temperatures is instead much closer to $T_{FT}^{q_E^R}$, with $T_{kin}^{dw}$ essentially coinciding exactly. We therefore conclude that it is not enough to limit our attention to the evolution of the particle around the potential minima, but rather considering its dynamics around its local maximum at $x_u$ is essential. Moreover, the affinity of $T_{FT}^{q_E^R}$ to the kinetic temperature seems to mirror the inherent character of $\mathcal{Q}_E^R$: instantaneous energy exchanges are best described in terms of an instantaneous out-of-equilibrium temperature.

\begin{figure}[t!]
\begin{center}
  \begin{tabular}{cc}
       \includegraphics[width=0.99\columnwidth]{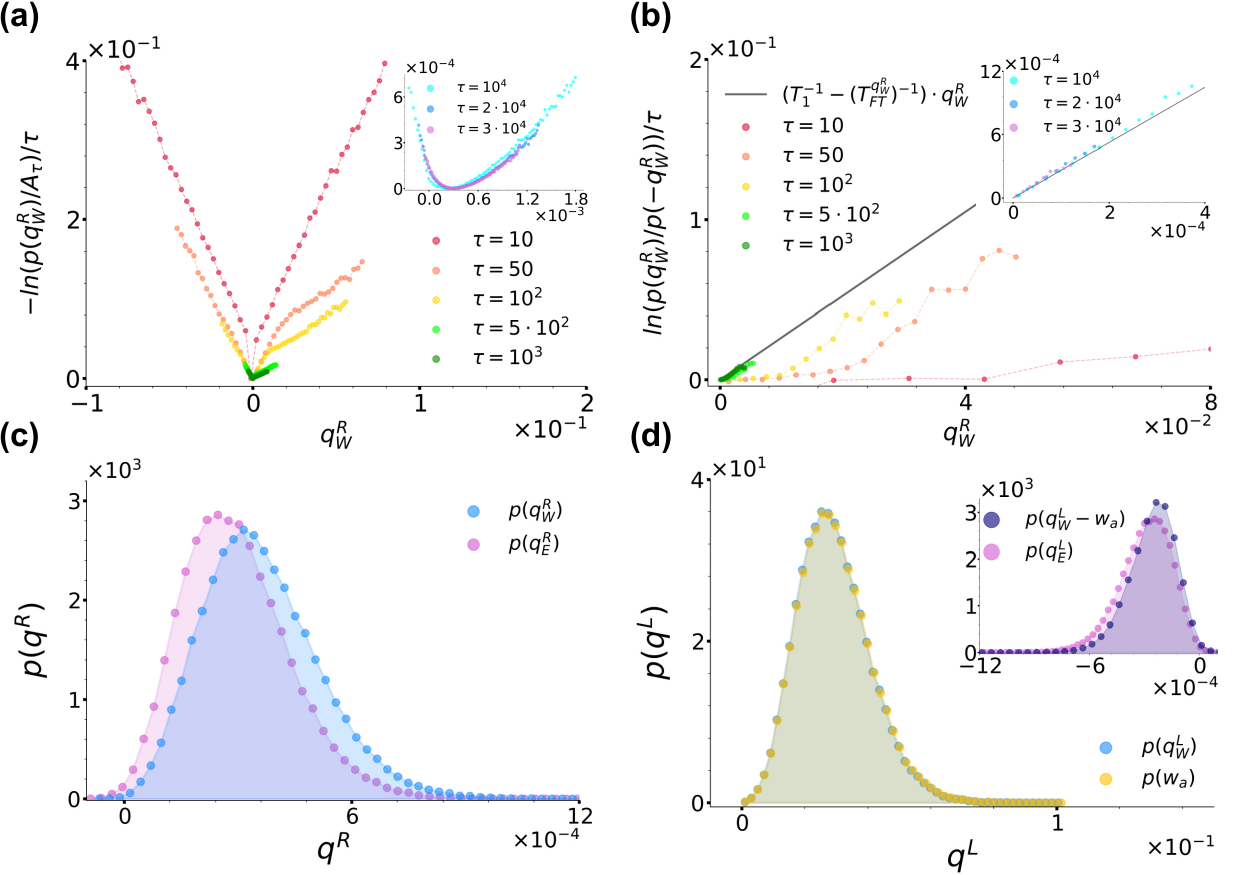}\quad
  \end{tabular}
\caption{\footnotesize{$\textbf{(a):}$ curves $-\ln(p(q_W^R))/\tau$ for $Pe=50$ at different sampling times, as denoted by the legend. $A_\tau$ denotes the maximum of the distribution at each sampling time. The inset shows instead the trend of the same curves at the largest sampling times considered. $\textbf{(b):}$ ratio $\ln(p(q_W^R)/p(-q_W^R))/\tau$ evaluated at different sampling times using data from panel $\textbf{(a)}$ along with the right hand-side of \autoref{eq:fr_qe_tr_tl} with $T_r=0.2$ fixed and $T_l=T_{FT}^{q_W^R}\sim 0.42$ extracted from a fit of the
curves in the inset performed as described in the main text, as denoted by the legend. $\textbf{(c)}$ and $\textbf{(d):}$ comparison between the distributions $p(q_E^R),~p(q_W^R)$ and $p(q_W^L),~p(w_a)$ at sampling time $\tau=3\cdot 10^4$, respectively, with same parameters as in \autoref{fig:active_baths}$\textbf{(a)}$. In panel $\textbf{(d)}$ the inset shows instead a comparison between $p(q_E^L)$ and $p(q_W^L-w_a)$. In all panels we fixed $\gamma=10$, $T_1=T_2=0.2$ and $a=1.0,~b=2.0$.}}
\label{fig:active_baths_bis}
\end{center}
\end{figure}

We now turn to comment on the behaviour of the heat $q_W$ and active work $w_a$ per unit time. \autoref{fig:active_baths_bis}$\textbf{(a)}$ reports the curves $-\ln(q_W^R)/\tau$ for increasing sampling time $\tau$ for the same parameter choice as in \autoref{fig:active_baths}$\textbf{(b)}$. Along with its inset, the figure shows that also in this case these curves converge at large times towards a convex rate function $I(q_W^R)$, thus proving $q_W^R$ to satisfy a large deviation principle also when the left bath is active. Concerning the validity of a fluctuation theorem, \autoref{fig:active_baths_bis}$\textbf{(b)}$ shows the ratio $\ln(p(q_W^R)/p(-q_W^R))/\tau$ evaluated at different times using data from \autoref{fig:active_baths_bis}$\textbf{(a)}$. As in \autoref{fig:passive_baths_bis}$\textbf{(d)}$, at small times the sub-exponential contribution $(c(q_W^R)-c(-q_W^R))/\tau$ makes the resulting curves actually curvilinear, while at large times they assume a linear trend with constant slope. Following the usual fitting procedure, in this case we find $T_{FT}^{q_W^R}\sim0.42$, and the resulting curve $(T_1^{-1}-(T_{FT}^{q_W^R})^{-1})\cdot q_W^R$ is reported in \autoref{fig:active_baths_bis}$\textbf{(b)}$ and its inset for completeness. In order to give context to this finding, \autoref{fig:active_baths_bis}$\textbf{(c)}$ reports a comparison between $p(q_W^R)$ and $p(q_E^R)$ at sampling time $\tau=3\cdot 10^4$. Interestingly, at variance with \autoref{fig:passive_baths_bis}$\textbf{(b)}$ from case a), here at large times the two distributions do not coincide, but rather $p(q_W^R)$ results slightly shifted towards the right with respect to $p(q_E^R)$. As \autoref{fig:active_baths}$\textbf{(b)}$ and \autoref{fig:active_baths_bis}$\textbf{(a)}$ show that at this $\tau$ the curves $-\ln(p(q_E^R))/\tau$ and $-\ln(p(q_W^R))/\tau$ have both already converged towards their respective rate functions, the origin of this discrepancy is not a matter of not long enough sampling time, but its explanation must rather be searched once again in the very dynamics of the system. To this end, we reconsider the position stationary distribution from \autoref{fig:model_pos_distr_app}$\textbf{(b)}$. As commented in \autoref{sec:stat_pos}, this is characterized by a left peak shifted towards the left with respect to the location of the left minimum $-x_m$ of the double-well potential \autoref{eq:potential}. This effect is in turn ascribable to the action of the active noise $a(t)$ which pushes the particle towards the left when assuming persistently negative values. When instead $a(t)$ persistently assumes positive values, it pushes the particle towards the right until making it jump in the right well. In doing so, the particle essentially takes a run-up, so that, at variance with case a), when jumping towards the right its velocity is enhanced. As a consequence the particle is not able to dissipate all accumulated excess energy essentially instantaneously at $x_u$ as in case a), but rather it completes its thermalization with the right well bath during the descent towards the right minimum located at $x_m$, hence the released surplus energy, the shift of $p(q_W^R)$ and the resulting fit temperature $T_{FT}^{q_W^R}$ higher than both $T_{FT}^{q_E^R},T^{dw}_{kin}\sim 0.3$. 

We remark that this finding is coherent with the phenomenology described above: the release of energy associated to $q_W^R$ does not occur instantaneously so the kinetic temperature is obviously not fit to describe this phenomenon. At the same this energy release does not persist long enough to make the effective temperature $T_{eff}^{dw}\sim21.2$ intervene, so that $T^{dw}_{kin}\sim T_{FT}^{q_E^R}<T_{FT}^{q_W^R}<T^{dw}_{eff}$. Interestingly, \autoref{fig:active_baths}$\textbf{(d)}$ and \autoref{tab:comparison_active} show that the temperature discrepancies we uncovered at $Pe=50$ are not peculiar of this specific case, but instead are common for all other $Pe$ we considered, with $T^{dw}_{kin}\sim T_{FT}^{q_E^R}<T_{FT}^{q_W^R}<T^{dw}_{eff}$ in all cases. 

To conclude, we briefly comment on the heat exchanges distributions for the left bath, captured at sampling time $\tau=3\cdot 10^4$ and $Pe=50$. \autoref{fig:active_baths_bis}$\textbf{(d)}$ reports $p(q_W^L)$ and $p(w_a)$ and shows that values spanned by these distributions are much larger than the ones spanned by $p(q_W^R)$ and $p(q_E^R)$ from \autoref{fig:active_baths_bis}$\textbf{(c)}$. This effect is due to the enhanced velocity of the particle pushed by the active force. Note also that the signs of $q_W^L$ and $w_a$ agree and their distributions almost overlap. That they do not completely overlap is in turn shown by the inset reporting $p(q_W^L-w_a)$ compared to $p(q_E^L)=p(-q_E^R)$. The inset in fact shows that $p(q_W^L-w_a)$ records non-zero values of three orders of magnitude lower than those associated to the distributions from the main figure and of the same order of magnitude as the ones associated to $p(q_E^L)$. Moreover, similarly to what happens for $p(q_W^R)$ and $p(q_E^R)$ from \autoref{fig:active_baths_bis}$\textbf{(c)}$, $p(q_W^L-w_a)$ is slightly shifted with respect to $p(q_E^L)$. Finally, we conclude by remarking that the distributions $p(q_W^R)$ and $p(q_W^L-w_a)$ result coherent with the energy balance \autoref{eq:en_bal_qe}.

\section{Conclusions}
\label{sec:conclusions}

In this paper we numerically studied the heat exchanges occurring between two heat baths of different nature with the purpose to investigate on the role temperature plays in these phenomena. The baths were spatially confined in the two wells of a quartic double well potential, while the heat exchanges were mediated by a Brownian particle jumping between the two. heat was sampled according to two different definitions: as the total kinetic energy carried by the particle when jump events occur and as the work performed by the particle on one of the two baths when immersed in it. The distributions of these heats were used to check the validity of a fluctuation theorem whence possibly extracting a temperature estimate for the baths through a proper linear fit. This procedure allowed us to introduce the definition of an out-of-equilibrium temperature whose resulting values we compared not only with the bath temperatures, but also with other out-of-equilibrium temperatures as the kinetic and effective ones. Operatively, we fixed an equilibrium bath in the right well and considered two different configurations for the one in the left well.

In the first case we fixed another equilibrium thermal bath with a different temperature and found both heats definitions to satisfy the same fluctuation theorem with fit temperatures coinciding with the ones of the two baths. These results extend the analysis of \cite{bodineau2007, visco2006, fogedby2011} to the cases of spatially separated equilibrium thermal baths.

In the second case we instead considered an active bath by introducing an additional Ornstein-Uhlenbeck noise, making the bath effectively out of equilibrium. Also in this case we found a fluctuation theorem to be satisfied. However, here the temperature relative to the left bath turns out to coincide with its out-of-equilibrium kinetic temperature for the heat defined as sum of kinetic energies and with an higher one still lower than the effective temperature for the other definition of heat. These results and discrepancies were interpreted by looking at the system phenomenology, finding them to mirror the instantaneous or longer release of energy captured by the the two heat definitions.

The present study could represent the first step towards a deeper and wider investigation on the role played by temperature in heat exchanges. If and where possible, analytical approaches could in fact provide further validation and insights to the overall scenario emerged from our investigation. Moreover more complex geometries and bath features could be explored so as to clarify even better the role of kinetic temperature in heat exchanges or  reveal cases where instead the effective temperature plays a prominent role. Experiments adopting setups and technologies already in place like Janus particles \cite{jiang2010, theurkauff2012, buttinoni2013, walther2008} and optical tweezers \cite{wang2002, blickle2006} could provide further validation and connections with real systems.

\vspace{6pt} 

\appendix

\section[\appendixname~\thesection]{Effective and Kinetic temperature}
\label{app:eff_kin_temp}

{\it Effective} and {\it kinetic} temperatures are two out of equilibrium temperature definitions able to capture instantaneous and time-delayed properties of the system \cite{Ilg2006, cugliandolo2011, loi2011, isaac2011, dieterich2015, levis2015, cugliandolo2019, nandi2018, mandal2019, petrelli2020}, respectively. Their definitions rely in fact on two fundamental results of statistical mechanics more effective in these two time regimes: the equipartition theorem \cite{alonso1967} and the fluctuation-dissipation theorem \cite{kubo1966}. 

Concerning the effective temperature, we first recall the definition of mean square displacement and integrated linear response function. The former is defined as
\begin{equation}
\Delta^2(t',t)\equiv\braket{[\bm{r}(t)-\bm{r}(t')]^2}~,
\end{equation}
and measures how far on average a particle travels over time with respect to a fixed initial location, while the latter is defined as
\begin{equation}
\chi(t',t)\equiv\int_{t'}^t dt'' \sum_{\alpha=1}^dR_{\alpha\alpha}(t'',t)~,
\end{equation}
with
\begin{equation}
R_{\alpha\beta}(t',t)=\frac{\delta\braket{r_\alpha(t)}_h^\lambda}{\delta h_\beta^\lambda(t')}\bigg|_{h_\beta^\lambda=0}
\end{equation}
linear response of the system, $d$ dimension of the system, $\alpha, \beta$ dimensional indices and $h_\beta^\lambda(t')$ an external perturbation depending on the parameter  $\lambda$ and measures how a system responds to a small external perturbation. The above two functions are both involved in the position fluctuation-dissipation theorem
\begin{equation}
2T\chi(t',t)=\Delta^2(t',t)~.
\label{eq:fdt_eq}
\end{equation}
Based on the above relation, the time dependent effective temperature for an out-of-equilibrium systems is then defined as \cite{cugliandolo2011, Ilg2006, cugliandolo2019}
\begin{equation}
T_{eff}(t',t)\equiv\frac{\Delta^2(t',t)}{2\chi(t',t)}~.
\label{eq:teff_def}
\end{equation}

Concerning the kinetic temperature, we recall the equipartition theorem to state that for each degree of freedom $i$ of an equilibrium system, the temperature $T$ and the velocity fluctuations are related as 
\begin{equation}
\frac{1}{2}m\Braket{\dot{\bm{r}}^2_i(t)}=\frac{1}{2}k_BT~.
\end{equation}
It is then natural for each degree of freedom of an out-of-equilibrium system to define the time-dependent kinetic temperature as \cite{Ilg2006, mandal2019, petrelli2020}
\begin{equation}
T_{kin}(t)\equiv \frac{m\braket{\dot{\bm{r}}^2_i(t)}}{k_B}~.
\label{eq:tkin_def}
\end{equation}

In equilibrium systems, as for example a single brownian particle in contact with a white-noise bath with temperature $T$, the two definitions boil down to the same expression, $T$. In more complex configurations they can instead be very different. For example, for a free active Ornstein-Uhlenbeck particle \cite{bonilla2019, caprini2021, cates2021, semeraro2021}, i.e. a particular instance of active particle, the integrated linear response function reads
\begin{equation}
\chi(t',t)=\frac{t-t'}{\gamma}~,
\end{equation}
while the mean square displacement and velocity expressions are provided by \cite{caprini2021}. Combining these functions as prescribed by \autoref{eq:teff_def} and \autoref{eq:tkin_def} and fixing $t'=0$ one finds that in the long time limit $t\uparrow\infty$
\begin{equation}
    T_{eff} \quad \longrightarrow\quad T+\frac{F_a^2}{\gamma\gamma_R}~.
    \label{eq:free_eff}
\end{equation}
and 
\begin{equation}
    T_{kin} \quad \longrightarrow\quad T+\frac{F_a^2}{\gamma}\frac{1}{(\gamma_R+\frac{\gamma}{m})}~,
    \label{eq:free_kin}
\end{equation}
where we dropped the time dependence as both temperatures reach a constant value. Analytical estimates can also be provided for an active Ornstein-Uhlenbeck particle under the action of an external harmonic potential like $U(x)=kx^2/2$. In this case the integrated linear response function is
\begin{equation}
\chi(t',t)=\frac{1-e^{-\frac{\gamma}{k}(t-t')}}{k}~,
\end{equation}
while the mean and velocity square displacement expressions are again provided by \cite{caprini2021}. Combining these functions as above and fixing $t'=0$ one finds that in the long time limit $t\uparrow\infty$
\begin{equation}
    T_{eff} \quad \longrightarrow\quad T+\frac{F_a^2}{\gamma}\frac{1+\frac{\gamma}{m\gamma_R}}{(\gamma_R+\frac{\gamma}{m}+\frac{k}{m \gamma_R})}~.
    \label{eq:t_eff_h}
\end{equation}
and
\begin{equation}
    T_{kin} \quad \longrightarrow\quad T+\frac{F_a^2}{\gamma}\frac{1}{(\gamma_R+\frac{\gamma}{m}+\frac{k}{m \gamma_R})}
    \label{eq:t_kin_h}
\end{equation}
which in the limit $k\downarrow 0$ reduce to \autoref{eq:free_eff} and \autoref{eq:free_kin}, respectively. 

Up to our knowledge, in even more complex systems, analytical results are not available. However, one can always resort to numerical methods and evaluate the kinetic and effective temperatures as prescribed by their definitions \autoref{eq:teff_def} and \autoref{eq:tkin_def}. In particular, the effective temperature estimation requires the knowledge of the integrated linear response function, to be evaluated using an external perturbation low enough for the system to remain in the linear regime and at the same time large enough to overcome large fluctuation effects.

\section*{References}
\bibliographystyle{iopart-num.bst}		
\bibliography{Tesi.bib}	        

\end{document}